\documentclass[twocolumn,aps,prb,superscriptaddress,longbibliography]{revtex4-2}
\usepackage{graphicx}
\usepackage{color}
\usepackage{amsmath}
\usepackage{enumitem}
\usepackage{amssymb}
\usepackage{hyperref}
\usepackage[normalem]{ulem}
\usepackage{adjustbox}

\hypersetup{
	colorlinks = true,
	linkcolor = blue,
	citecolor = blue,
	urlcolor  = blue,
}

\newcommand{\be}{\begin{equation}}
	\newcommand{\ee}{\end{equation}}

\newcommand{\bea}{\begin{eqnarray}}
	\newcommand{\eea}{\end{eqnarray}}

\newcommand{\p}{\partial}

\newcommand{\la}{\left\langle}
\newcommand{\ra}{\right\rangle}

\newcommand{\sgn}{{\rm sgn\,}}
\newcommand{\Tr}{{\rm \, Tr\,}}
\newcommand{\tr}{{\rm \, tr\,}}
\renewcommand{\Re}{{\rm \, Re\,}}
\renewcommand{\Im}{{\rm \, Im\,}}

\renewcommand{\vec}[1]{{\boldsymbol #1}}
\renewcommand{\epsilon}{\varepsilon}

\def\nn{\nonumber\\}

\newcommand{\addDO}[1]{\textcolor{blue}{#1}}

\begin{document}
	\title{Optical conductivity of semi-Dirac and pseudospin-1 models: Zitterbewegung  approach}
	\date{\today}

	\author{D. O. Oriekhov}
	\affiliation{Instituut-Lorentz, Universiteit Leiden, P.O. Box 9506, 2300 RA Leiden, The Netherlands}
	
	\author{V. P. Gusynin}
	\affiliation{Bogolyubov Institute for Theoretical Physics, National Academy of Science of Ukraine, 14-b Metrologichna Street, Kyiv, 03143, Ukraine}
	
	\begin{abstract}
	    We present a method to calculate the optical conductivity of semi-Dirac and pseudospin models based on the evaluation of quasiparticle velocity correlators which also describe the phenomenon of zitterbewegung. Applying this method to the semi-Dirac model with merging Dirac cones and gapped dice and Lieb lattice models we find exact analytical expressions for optical longitudinal and Hall conductivities. For the semi-Dirac model the obtained expressions allow us to analyze the role of spectrum anisotropy, van Hove singularities and Dirac cones in longitudinal conductivity. In addition, we predict signatures of topological phase transition with changing gap parameter in such a system that are manifested in dc transport at low temperatures. For the dice and Lieb lattices we emphasize the role of spectral gap, which defines frequency thresholds related to transitions to and from flat band.
	\end{abstract}
	\maketitle
	
	\section{Introduction}
The optical studies of electronic systems is one of the main sources of information  about charge dynamics in different condensed matter systems: high-Tc superconducting cuprates \cite{Basov2005,Carbotte2011}, graphene \cite{Ando2002,Gusynin2006PRB,Gusynin2006PRL,Nair2008Science,Li2008Nature,Stauber2008PRB}, topological insulators \cite{Schafgans2012} together with Dirac and Weyl materials \cite{Chen2015,Xu2016PRB,Neubauer2016PRBR}. 	
	Recently it was shown \cite{Bradlyn2016} that in crystals with special space symmetry groups more complicated quasiparticle spectra could be realized
with no analogues in high-energy physics where the Poincare symmetry provides strong restrictions. Some of such systems possess strictly flat (dispersionless) bands \cite{Heikkilae2011,Heikkilae2011a,Leykam2018} with high degeneracy potentially leading to a large enhancement of some physical quantities.

In the present paper we develop the method to calculate frequency-dependent optical and Hall conductivities in low-energy models containing also new types of quasiparticles. The presented method is based on the solution of the Heisenberg equations for the time-dependent quasiparticle velocity operators, which also describe the phenomena of zitterbewegung (trembling motion) \cite{Katsnelson2006EPJB,Cserti2006PRB}. The formulation of this method is very similar to the proper time approach discussed by Schwinger \cite{Schwinger1951} and the obtained expression extend previously derived formulas for longitudinal conductivity in Refs.\cite{Gusynin2007conductivity,Yuan2010PRB}. We  rewrite the Kubo formula \cite{Kubo1957} through quasiparticle velocity correlators, and use the solutions of the Heisenberg equations. We demonstrate the applicability of the described method to the semi-Dirac model and gapped pseudospin-1 models of the dice and Lieb lattices. As a result, we obtain closed-form analytic expressions, which in turn are used to investigate the dependence of conductivity on frequency, gap size and temperature.

The phenomenon of Dirac points merging in two-dimensional materials has received much attention in the literature \cite{Hasegawa2006PRB,Montambaux2009PRB,Montambaux2009EPJ}. Such system was realized experimentally in optical lattices \cite{Tarruell2012Nature} and in microwave cavities \cite{Bellec2013PRL}. The analytical and numerical
calculations of optical conductivity for  semi-Dirac systems were discussed in several recent papers \cite{Adroguer2016,Ziegler2017,Mawrie2019,Carbotte2019,Carbotte2019PRB,Jang2019,Carbotte2019,Carbotte2019PRB}. Quite recently the
magneto-conductivity of the semi-Dirac model was studied \cite{Zhou2021PRB}.

	The dice model is a tight-binding model of two-dimensional fermions living on the ${\cal T}_3$ (or dice) lattice where atoms are situated both at the vertices of hexagonal lattice and the hexagons centers \cite{Sutherland1986,Vidal1998}. Since the dice model has three sites per unit cell, the electron
states in this model are described by three-component fermions and the energy spectrum of the model is comprised of three bands. The two of them form
	 Dirac cones and the third band is completely flat and has zero energy \cite{Bercioux2009,Raoux2014}.  The ${\cal T}_3$ lattice has been experimentally realized in Josephson arrays \cite{Serret2002,Abilio1999}, metallic wire networks \cite{Naud2001} and its optical realization by laser beams was proposed in Refs.\cite{Rizzi2006,Bercioux2009}.
	The optical and Hall conductivities for the $\alpha-\mathcal{T}_3$ model were studied in Refs. \cite{Illes2015,Chen2019,Dey2020,Han2022PRB}.  We show that our method allows one to obtain fully analytic expressions for the case of $S_z$ model even without magnetic field, thus extending the previous results.
	
	Another example of pseudospin-1 system considered in this paper is the gapped low-energy model of the Lieb lattice \cite{Shen2010}. Due to the presence of flat band in spectrum \cite{Shen2010,Apaja2010PRA,Goldman2011}, the Lieb lattice served as a platform for theoretical studies of many strongly-correlated phenomena - ferromagnetism \cite{Lieb1989PRL,Tasaki1998} and superconductivity \cite{Kopnin2011PRB,Julku2016PRL}.
The Lieb lattice was realized in many experimental setups: arrays of optical waveguides \cite{Vicencio2015PRL,Mukherjee2015},  in  vacancy lattice in chlorine monolayer on  Cu(100) surface \cite{Drost2017Nature} and in covalent organic frameworks \cite{Jiang2019,Cui2020Nature}.
	
	The paper is organized as follows: in Sec.\ref{sec:conductivity-general} we present the most general formulas for the optical and Hall conductivity in terms of quasiparticle velocity correlators. In Sec.\ref{sec:semi-Dirac} we apply the method for a simple, but physically reach semi-Dirac model with merging Diral cones.  Next, we apply the described approach to calculate the optical conductivity of the gapped dice model. For this purpose in Sec.\ref{sec:dice-Heisenberg} we solve the Heisenberg equations for the dice model with gap and discuss properties of the quasiparticle dynamics. Combining the results with general formulas for conductivity in Sec.\ref{sec:conductivity}, we find the optical and Hall conductivity and analyze their dependence on external frequency. Finally, in Sec.\ref{sec:Lieb-optical} we perform similar calculation for the Lieb lattice model, whose underlying matrix algebra is much more complicated. In the Appendices we present the details of Kubo formula transformations and conductivity integrals evaluation.
	
	\begin{figure*}
		\centering
		\includegraphics[scale=0.5]{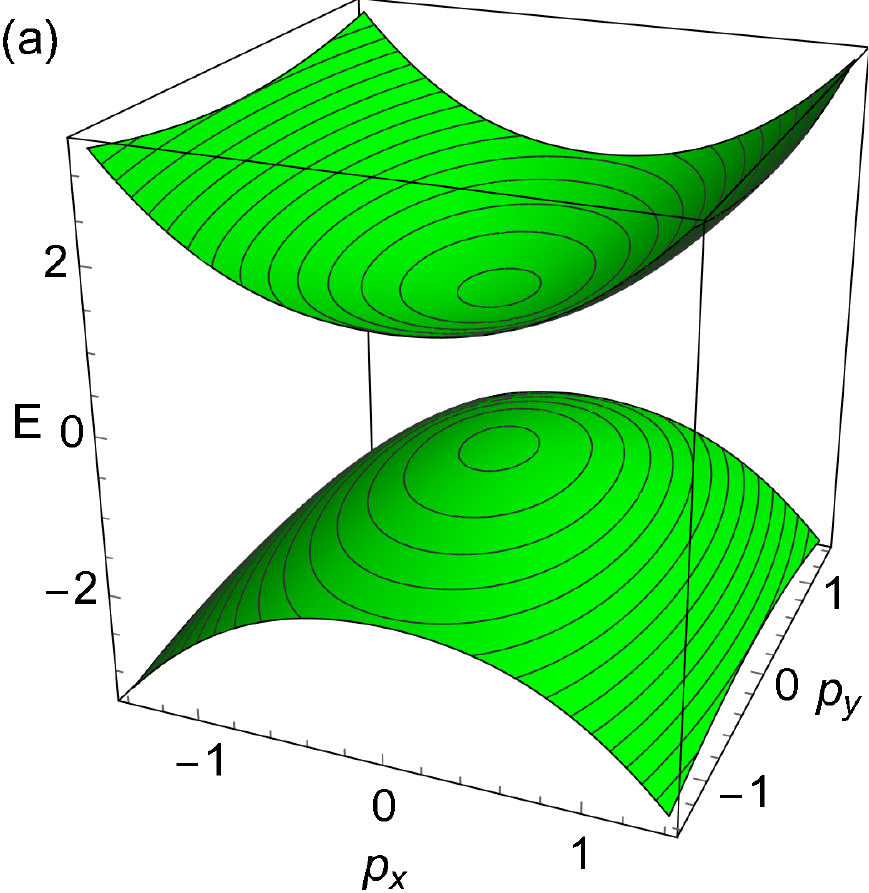}\qquad
		\includegraphics[scale=0.5]{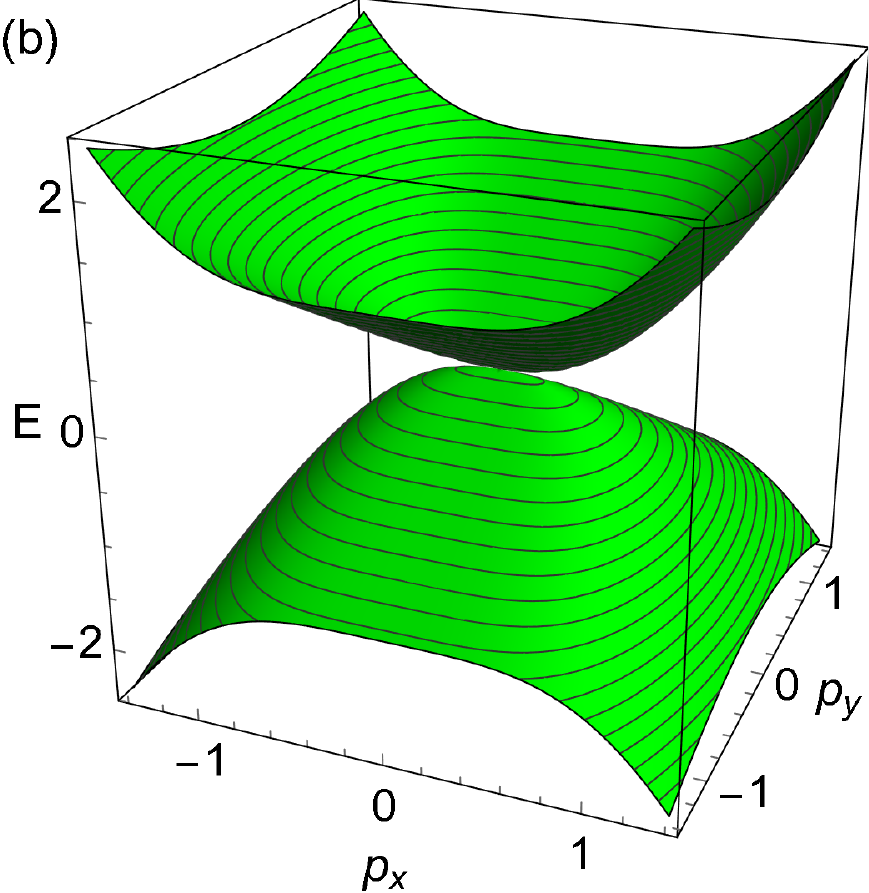}\qquad
		\includegraphics[scale=0.5]{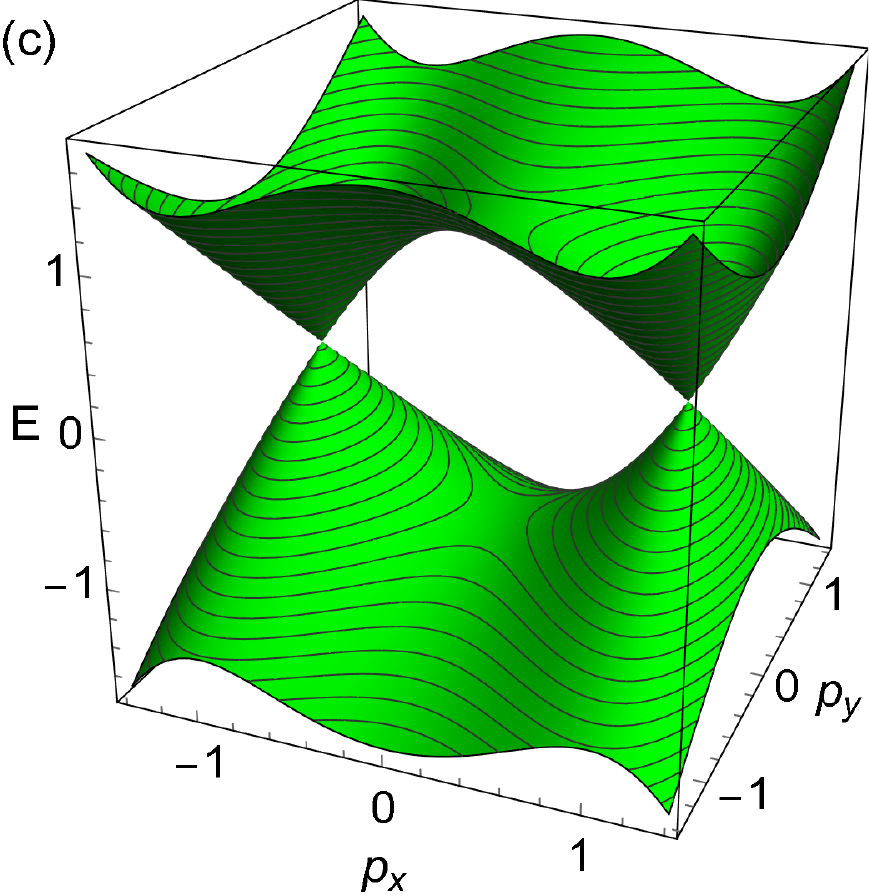}
		\caption{Spectrum given by Hamiltonian $H_{semi}$ in Eq.\eqref{eq:semi-1}. The values of gap parameter are (a) $\Delta=1$, (b) $\Delta=0$ and (c) $\Delta=-1$. We choose units $v=1,\,a=1$. The panel (a) represents a a fully gapped regime, while the panel (c) corresponds to the regime with two Dirac cones separated by $2\sqrt{\Delta/a}$ along the x-direction.}
		\label{fig:semi-spectrum}
	\end{figure*}
	
	\section{Expression for conductivity through particle velocity correlators}
	\label{sec:conductivity-general}
	The method described below is an extension of the approach used in Ref.\cite{Katsnelson2006EPJB} to an arbitrary pseudospin model with different dispersions. We start the derivation from the Kubo formula \cite{Kubo1957} for frequency-dependent electrical conductivity tensor written in the following form \cite{Yuan2010PRB}:
\begin{align}
		&\sigma_{\mu \nu}(\omega)=\frac{i}{(\omega+i \epsilon) V}\nn
		&\times\left[\langle\tau_{\mu \nu}\rangle -\frac{i}{\hbar} \int_{0}^{\infty} d t e^{i(\omega+i \epsilon)t}\Tr\left(\hat{\rho}
\left[J_{\mu}(t), J_{\nu}(0)\right]\right)\right],
\label{Kubo-formula}
	\end{align}
where $V$ is the volume (area) of the system, $\hat{\rho}=\exp\left(-\beta H\right)/Z$ is the density matrix with
the Hamiltonian $H$ in the grand canonical ensemble, $Z=\Tr\exp\left(-\beta H\right)$ is the partition function, $\beta=1/k_BT$, and $J_\mu$ are
the current operators.
The diamagnetic or stress tensor $\langle\tau_{\mu \nu}\rangle$ in the Kubo formula (\ref{Kubo-formula}) is a thermal average of the operator
 defined as $\tau_{\mu \nu}=\partial^2H/\partial(A^\mu/c)\partial(A^\nu/c)$. In the case of a linear dispersion law the term with $\la\tau_{\mu\nu}\ra$ in Eq.(\ref{Kubo-formula}) is absent. In what follows we set $\hbar=1$ and restore it in the final expressions.

The important symmetry properties of the conductivity are
	\begin{align}
		\operatorname{Re} \sigma_{\mu \nu}(\omega) &=\operatorname{Re} \sigma_{\mu \nu}(-\omega), \\
		\operatorname{Im} \sigma_{\mu \nu}(\omega) &=-\operatorname{Im} \sigma_{\mu \nu}(-\omega).
	\end{align}
Using the representation of conductivity tensor through the correlation functions of currents (see Ref.\cite{Gusynin2007conductivity} and Appendix \ref{appendix:A}) and expressing them in terms of time-dependent particle velocity  correlators, we arrive at the following general expressions:
	\begin{align}\label{eq:Re-sigma-mu-nu}
		&\operatorname{Re} \sigma_{\{\mu, \nu\}}(\omega)=\frac{e^{2}}{2 
			 \omega} \int_{-\infty}^{\infty} d E \rho(E)\left[f(E)-f(E+
		  \omega)\right]
\nn &\times\int_{-\infty}^{\infty} d t e^{i \omega t}\left\langle v_{\{\mu}(t) v_{\nu\}}(0)\right\rangle_{E},
	\end{align}
where the velocity operator $v_\mu(t)=e^{iHt}v_\mu(0)e^{-iHt}$. Here we defined the microcanonical average of an operator $\hat{A}$
at given energy $E$ as
    \begin{equation}\label{eq:fixed-E-average}
    	\langle\hat{A}\rangle_{E}=\frac{\operatorname{Tr}[\delta(E-\hat{H}) \hat{A}]}{\operatorname{Tr}[\delta(E-\hat{H})]}
    \end{equation}
    where $\operatorname{Tr}[\delta(E-\hat{H})]=\rho(E) V$ and $\rho(E) $ is the density of states (DOS).
	It is easy to check that the last expression is real using
	\begin{equation}
		\left\langle v_{\{\mu}(-t) v_{\nu\}}(0)\right\rangle_{E}^{*}=\left\langle v_{\{\mu}(t) v_{\nu\}}(0)\right\rangle_{E}.
	\end{equation}
	The expression \eqref{eq:Re-sigma-mu-nu} for $T=0$ is in accordance with Ref.\cite{Mayou2000PRL} for diagonal conductivity.
	The numerator in Eq.\eqref{eq:fixed-E-average} can be represented using the Fourier transformation:
	\begin{align}\label{eq:trace-Fourier-sE-general}
		&\operatorname{Tr}[\delta(E-\hat{H}) \hat{A}]=\frac{V}{2\pi}\int_{-\infty}^{\infty} d s e^{i E s}\Tr[e^{-i\hat{H}s}\hat{A}]\nn
		&=\frac{V}{2\pi}\int_{-\infty}^{\infty} d s e^{i E s}\int\frac{d^2p}{(2\pi)^2}\tr[e^{-iH(\vec{p})s}\hat{A}(\vec{p})].
	\end{align}
Similarly, for the imaginary antisymmetric part of conductivity we have
	\begin{align}\label{eq:Im-sigma-mu-nu}
		&\operatorname{Im} \sigma_{[\mu, \nu]}(\omega)=\frac{e^{2}}{2\omega} \operatorname{Im} \int_{-\infty}^{\infty} d E \rho(E)\left[f(E)-f(E+\hbar \omega)\right] \nn
		&\times\int_{-\infty}^{\infty} d t e^{i \omega t}\left\langle v_{[\mu}(t) v_{\nu]}(0)\right\rangle_{E}.
	\end{align}
We note that the integral over $t$ is purely imaginary due to the property $\left\langle v_{[\mu}(-t) v_{\nu]}(0)\right\rangle^*_{E}
=-\left\langle v_{[\mu}(t) v_{\nu]}(0)\right\rangle_{E}$.
	
To calculate $\operatorname{Im} \sigma_{\{\mu, \nu\}}(\omega)$ and $\operatorname{Re} \sigma_{[\mu, \nu]}(\omega)$ we use the Kramers-Kr\"{o}nig relation \eqref{eq:Kramers-Kronig}. The equations \eqref{eq:Re-sigma-mu-nu} and \eqref{eq:Im-sigma-mu-nu} together with Eqs.\eqref{eq:fixed-E-average} and \eqref{eq:trace-Fourier-sE-general} allow one to obtain the final result after two Fourier transformations.
	
	\section{Optical conductivity of the semi-Dirac model}
	\label{sec:semi-Dirac}
	In this section we analyze the conductivity of the semi-Dirac model, which was extensively used to describe the low-energy physics of phosphorene \cite{Ezawa2015,Pyatkovskiy2016PRB,Carbotte2019,Carbotte2019PRB,Adroguer2016,Jang2019}. The main feature of such model is that it mixes linear and quadratic terms in the Hamiltonian
	\begin{align}\label{eq:semi-1}
		H_{semi}=\left(\Delta+a p_{x}^{2}\right) \sigma_{x}+v p_{y} \sigma_{y}.
	\end{align}
	The dispersion defined by this Hamiltonian consists of two bands:
	\begin{align}\label{eq:epsilon-semi-pm}
		\epsilon_{\pm}=\pm\sqrt{\left(a p_x^2+\Delta \right){}^2+v^2 p_y^2}.
	\end{align}
	The spectrum described by Eq.\eqref{eq:epsilon-semi-pm} is presented in Fig.\ref{fig:semi-spectrum}. By tuning the gap parameters, one can achieve a completely different types of spectrum - fully gapped, one band-touching point or two band-touching points separated by $2\sqrt{\Delta/a}$ distance along
$p_x$ momentum.
	
Writing the Heisenberg equations for this Hamiltonian, we find
	\begin{align}\label{eq:Heisenberg-semi-1}
		&\vec{v}(t)=\frac{d \vec{x}}{d t}=-i 
		 [ \vec{x}(t),\,H_{semi}(t)]=(2 a p_x(t) \sigma_x(t),\, v \sigma_y(t)),\\
		\label{eq:Heisenberg-semi-2}
		&\frac{d p_{i}}{d t}=- i 
		 [p_{i}, \,H_{semi}]=0.
	\end{align}
From the first equation we find that velocity depends on momentum $p_x(t)$, which does not evolve as a result of the second equation: $p_x(t)=p_x(0)$. Also,
velocity depends on the Pauli matrices, which evolve with time according to another Heisenberg equation:
	\begin{align}
	\frac{d\vec{\sigma}(t)}{dt}=-i[\vec{\sigma}(t),H_{semi}]=2[\tilde{\vec{p}}(0)\times \vec{\sigma}(t)].
	\end{align}
Here we used notation $\tilde{\vec{p}}(0)=[\Delta+a p_{x}^{2},v p_{y}, 0]$ and the fact that the commutator of the Pauli matrices is $[\sigma_i,\sigma_j]=2i \epsilon_{ijk}\sigma_k$. Cross means the vector product of $\tilde{\vec{p}}$ and $\vec{\sigma}$.  The initial condition for the Pauli matrices is $\vec{\sigma}(0)=(\sigma_x,\, \sigma_y,\,\sigma_z)$. The Heisenberg equation above gives a system of differential equations for matrices $\dot{\sigma}_i(t)=P_{ij} \sigma_j(t)$, $P_{ij}=2\epsilon_{ikj}\tilde{p}_k$, whose solution is
	\begin{align}\label{eq:solution-matrix-semi-Dirac}
		&\sigma_i(t)=\left(e^{P t}\right)_{i j}(\tilde{\boldsymbol{p}})\sigma_j(0),\quad\quad \left(e^{P t}\right)_{i j}(\tilde{\boldsymbol{p}})=\nn
		&\left(\begin{array}{ccc}
\frac{\tilde{p}_{y}^{2} \cos (2 \tilde{p} t)+\tilde{p}_{x}^{2}}{\tilde{p}^{2}} & \frac{\tilde{p}_{x} \tilde{p}_{y}(1-\cos (2 \tilde{p} t))}{\tilde{p}^{2}} & \frac{\tilde{p}_{y} \sin (2 \tilde{p} t)}{\tilde{p}} \\
			\frac{\tilde{p}_{x} \tilde{p}_{y}(1-\cos (2 \tilde{p} t))}{\tilde{p}^{2}} & \frac{\tilde{p}_{x}^{2} \cos (2 \tilde{p} t)+\tilde{p}_{y}^{2}}{\tilde{p}^{2}} & -\frac{\tilde{p}_{x} \sin (2 \tilde{p} t)}{\tilde{p}} \\
			-\frac{\tilde{p}_{y} \sin (2 \tilde{p} t)}{\tilde{p}} & \frac{\tilde{p}_{x} \sin (2 \tilde{p} t)}{\tilde{p}} & \cos (2 \tilde{p} t)
		\end{array}\right).
	\end{align}
Here we denoted $\tilde{p}=\sqrt{\tilde{p}_{x}^{2}+\tilde{p}_{y}^{2}}$. The time-dependent velocity is obtained from these solutions by combining
them with Eq.\eqref{eq:Heisenberg-semi-1}. The velocity $v_i(t)$ contains zitterbewegung terms which stem from the oscillatory terms (the cosine
and sine terms) in Eq.(\ref{eq:solution-matrix-semi-Dirac}).

Next we calculate the traces of velocity products with matrix exponential of the Hamiltonian as they appear in Eq.\eqref{eq:trace-Fourier-sE-general}.
 Due to the anisotropy in the electron dispersion, the conductivity is also anisotropic, therefore, we present the
results of its calculation in separate sections.

\subsection{Optical conductivity in xx-direction}

We start with the evaluation of real part of optical conductivity in the x-direction. For this purpose we start with the calculation of trace which has the form as in Eq.\eqref{eq:trace-Fourier-sE-general}:
\begin{align}\label{eq:tr-xx-semi-v2}
	&\Tr[e^{-i H_{semi}s}v_x(t)v_x(0)]=\int \frac{d^2 p}{(2\pi)^2} \frac{8 a^2 p_x^2}{\epsilon_{+}^2}\times\nn
	&\bigg(v^2 p_y^2 \cos \left((s-2 t)\epsilon_{+}\right)+\left(a p_x^2+\Delta
	\right){}^2 \cos \left(s \epsilon_{+}\right)\bigg).
\end{align}
Next we substitute this result into the expression for the real part of the xx longitudinal conductivity \eqref{eq:Re-sigma-mu-nu}, and calculate the Fourier transforms over $t$ and $s$. The result has the form of double integral:
\begin{align}
	&\Re\sigma_{xx}(\omega)=\frac{e^2}{ \omega}\int_{-\infty}^{\infty}\frac{d E}{2\pi} [f(E)-f(E+\omega)] \int d^2 p \frac{2 a^2 p_x^2}{\epsilon_{+}^2}\nn
	&\times\bigg[ \delta \left(E+\epsilon_{+}\right) \left(v^2 p_y^2 \delta \left(\omega
		+2\epsilon_{-}\right)+\delta
		(\omega ) \left(a p_x^2+\Delta \right)^2\right)\bigg.\nn
	&+\bigg.\delta
		\left(E+\epsilon_{-}\right)
		\left(v^2 p_y^2 \delta \left(\omega +2\epsilon_{+}\right)+\delta (\omega ) \left(a p_x^2+\Delta
		\right)^2\right)\bigg].
\end{align}

\begin{figure*}
	\centering
	\includegraphics[scale=0.33]{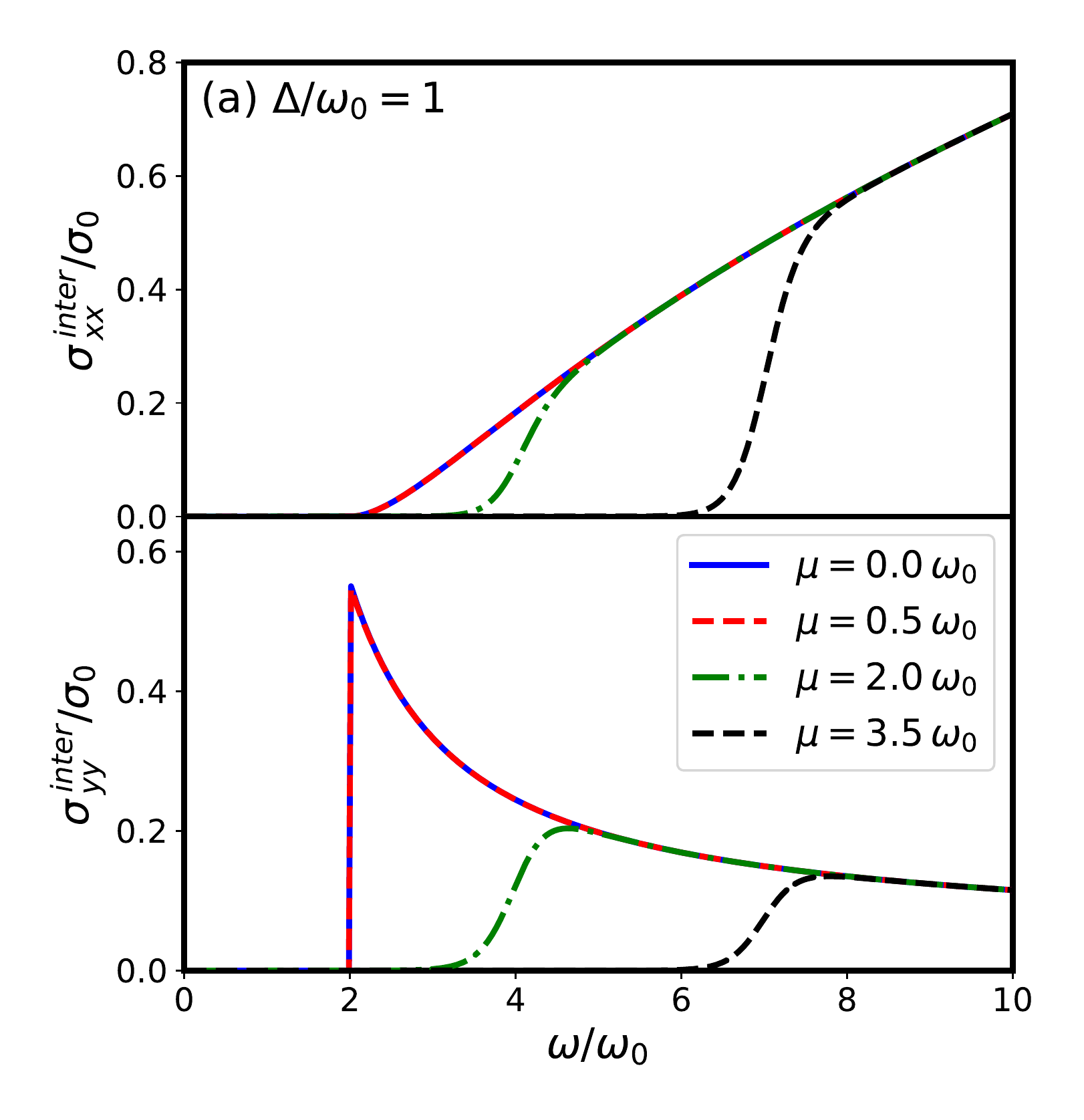}
	\includegraphics[scale=0.33]{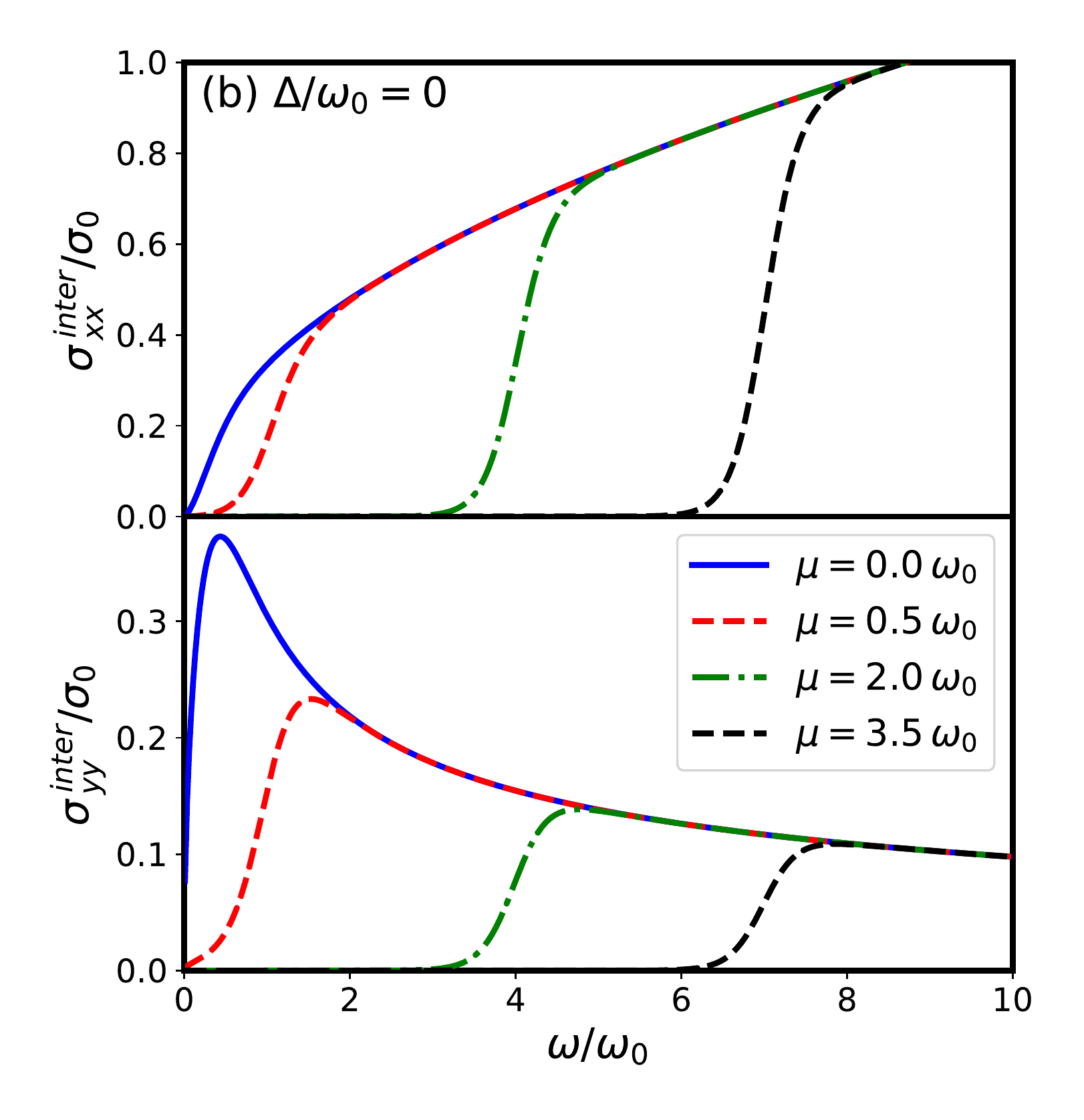}
	\includegraphics[scale=0.33]{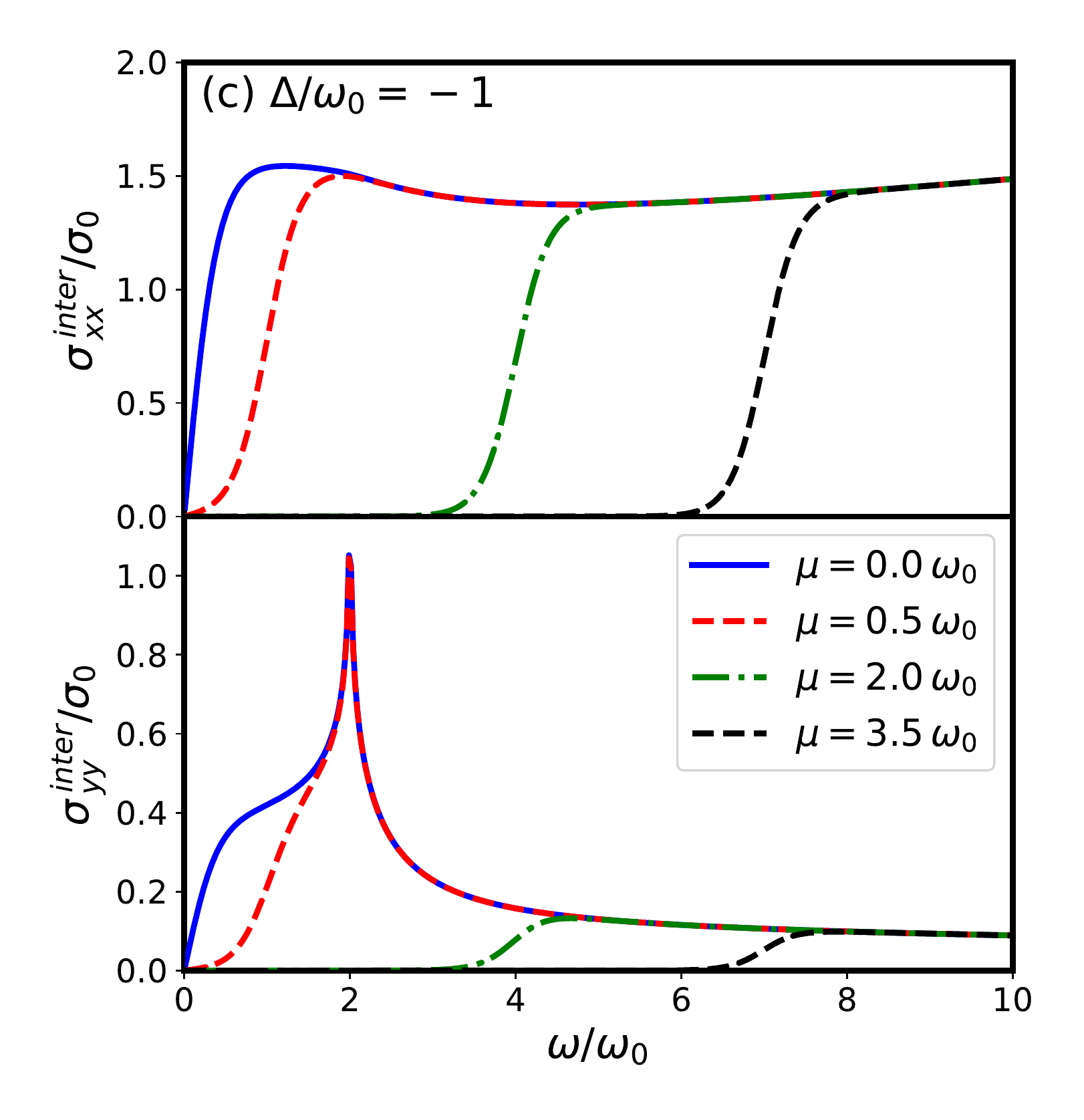}
	\caption{Real part of longitudinal interband ac conductivity in x- and y-directions (top and bottom plots) as a function of frequency for the fixed values of gap $\Delta$ for the semi-Dirac model. The frequency is measured in units of $\omega_0=v^2/a$. The normalization parameters are $\sigma_0=\frac{e^2 \sqrt{a}}{2\pi\hbar v}$ for the x-direction and $\sigma_0=\frac{e^2 v}{2\pi\hbar \sqrt{a}}$ for the y-direction. The values of gap parameter are (a) $\Delta /\omega_0 = 1$, (b) $\Delta /\omega_0=0$ and (c) $\Delta /\omega_0=-1$.}
	\label{fig:sigma-semi-AC-w-dep}
\end{figure*}
 The procedure of integration over momentum depends on the sign of $\Delta$ parameter, and is described in details in Appendix \ref{appendix:conductivity-semi-Dirac-xx-v2}. The main trick in calculation is to introduce modified polar coordinates, which take into account the anisotropy of dispersion \eqref{eq:epsilon-semi-pm} in each case $\Delta<0$, $\Delta=0$ and $\Delta>0$ with the proper regions of integration. As a result, we were able to express all integrals in terms of complete elliptic integrals. The results for the real part of interband ac and intraband dc conductivities are:
 	\begin{align}\label{eq:Re-sigma-semi-xx-AC}
 		&\operatorname{Re} \sigma_{x x}^{inter}(\omega)=\sgn\omega \frac{e^{2}}{2\pi \hbar} \frac{\sqrt{2|\omega|a}}{ 4 v}\left[f\left(-\frac{\omega}{2}\right)-f\left(\frac{\omega}{2} \right)\right] \times\nn
 		&\times\begin{cases}
 			\begin{array}{c}	
 			2\Theta(|\Delta|-|\omega/2|) I_{3}^{xx}(2\Delta/ |\omega|)\\
 			+2\Theta(|\omega/2|-|\Delta|) I_{1}^{xx}(2\Delta/|\omega|)
 			\end{array}
 		,&\Delta<0,\\
 		\\
 			\frac{16\pi^{3 / 2}}{5 \sqrt{2} \Gamma^{2}\left(\frac{1}{4}\right)},&\Delta=0,\\
 			\\
 			2\Theta(|\omega/2|-\Delta)I_{1}^{xx}(2\Delta/|
 			\omega|),&\Delta>0.	
 		\end{cases}
 	\end{align}
 The integrals $I_{1}^{xx},\,I_{3}^{xx}$, and similar integrals occurring below,  are defined in Appendix \ref{appendix:conductivity-semi-Dirac-xx-v2}, they are given in terms of complete elliptic integrals of the first and second kind.

We plot the conductivity $\operatorname{Re} \sigma_{x x}^{inter}(\omega)$ as a function of $\omega$ at different values of $\Delta$  in upper plots of Fig.\ref{fig:sigma-semi-AC-w-dep}. In all plots we set $Ta=0.1$, and absorb $v$ and $a$ parameters into normalization constant $\sigma_0$. As is seen, the  behavior of the
conductivities at small frequencies, $\omega<2|\Delta|$, is radically different for $\Delta>0$ and $\Delta<0$: the case $\Delta>0$ corresponds to insulating
phase while $\Delta\leq0$ corresponds to metallic phase.

The analytic expression (\ref{eq:Re-sigma-semi-xx-AC}) allows one to get asymptotes at small and large $\omega$, for example, in the most
interesting case $\Delta<0$ they are
\bea
\operatorname{Re} \sigma_{x x}^{inter}(\omega)\simeq \frac{e^{2}}{2\pi\hbar}\left\{\begin{array}{cc}\frac{\sqrt{|\Delta|a}}{v}\frac{\pi\omega}{8T\cosh^2\frac{\mu}{2T}},
\quad \omega\to0,\\
\frac{\sqrt{\omega a}}{v}\frac{4\pi^{3/2}}{5\Gamma^2\left(\frac{1}{4}\right)},\quad\omega\to\infty.\end{array}\right.
\eea

In the intraband part of conductivity with $\delta(\omega)$ the result contains integral over energy,
 	\begin{align}\label{eq:Re-sigma-semi-xx-DC}
 		&\operatorname{Re} \sigma_{x x}^{intra}(\omega)=\delta(\omega) \frac{e^{2}\sqrt{a}}{4\pi \hbar vT} \int_{-\infty}^{\infty} \frac{dE|E|^{3 / 2}}{\cosh^2\left(\frac{E-\mu}{2T}\right)} \times\nn
 		&\times\begin{cases}
 				\begin{array}{c}
 			2\Theta(|\Delta|-|E|) I_{4}^{xx}(\Delta/ |E|) \\
 			+2\Theta(|E|-|\Delta|) I_{2}^{xx}(\Delta/|E|)
 			\end{array},&\Delta<0,\\
 		\\
 			\frac{3\pi^{3 / 2}}{10 \sqrt{2} \Gamma^{2}\left(\frac{5}{4}\right)},&\Delta=0,\\
 			\\
 			2\Theta(|E|-\Delta)I_{2}^{xx}(\Delta/|
 			E|),&\Delta>0.	
 		\end{cases}
 	\end{align}
The integral over energy can be evaluated  analytically only in the special case of zero temperature  $T\to0$. We plot $\operatorname{Re} \sigma_{x x}^{intra}$ as a function of the gap parameter $\Delta$ in Fig.\ref{fig:sigma-semi-DC-Delta-dep}. One can observe the monotonous decrease with growing $\Delta$ for all values of chemical potential.

\subsection{Optical conductivity in the y-direction}
	\begin{figure*}
	\centering
	\includegraphics[scale=0.35]{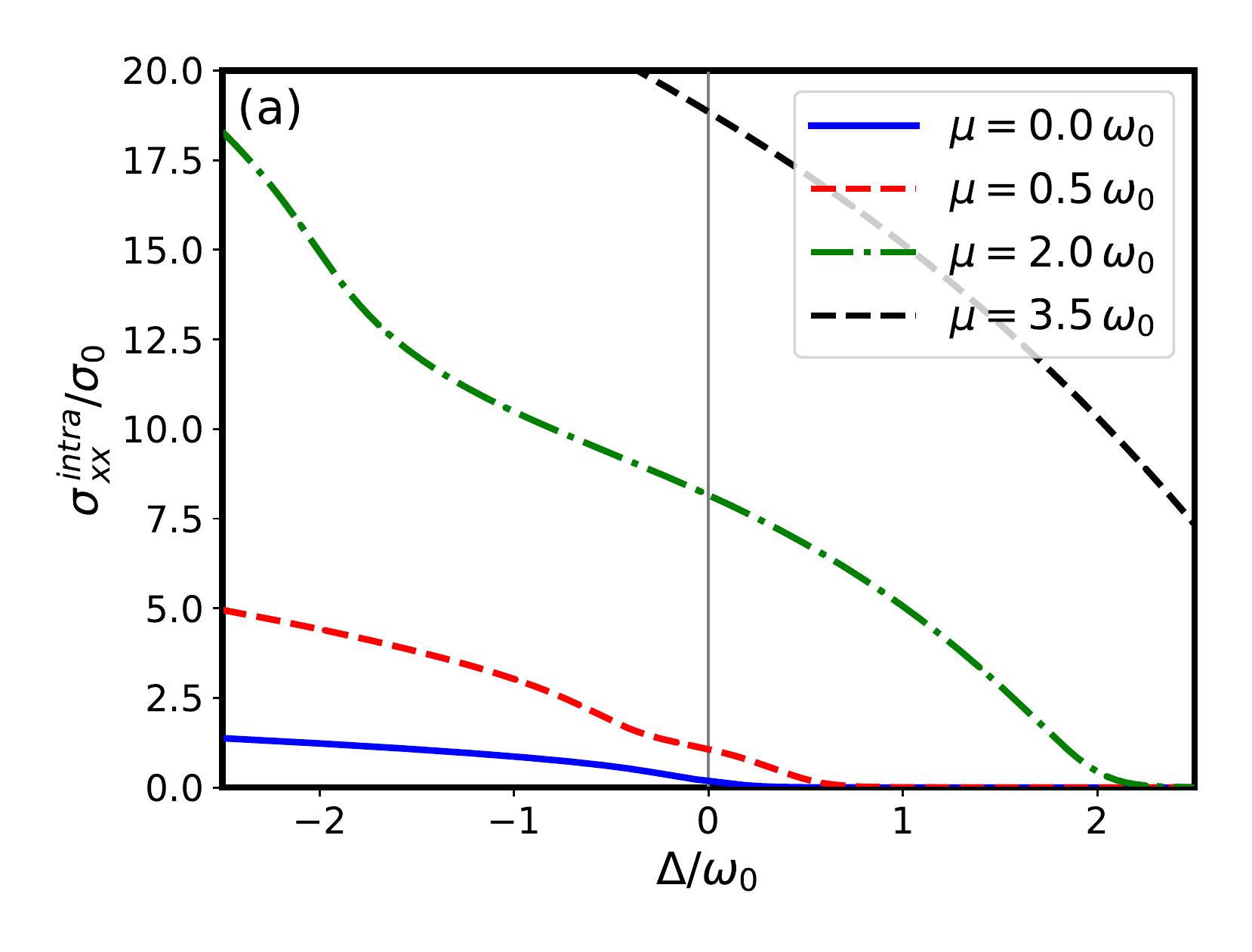}\qquad\qquad
	\includegraphics[scale=0.35]{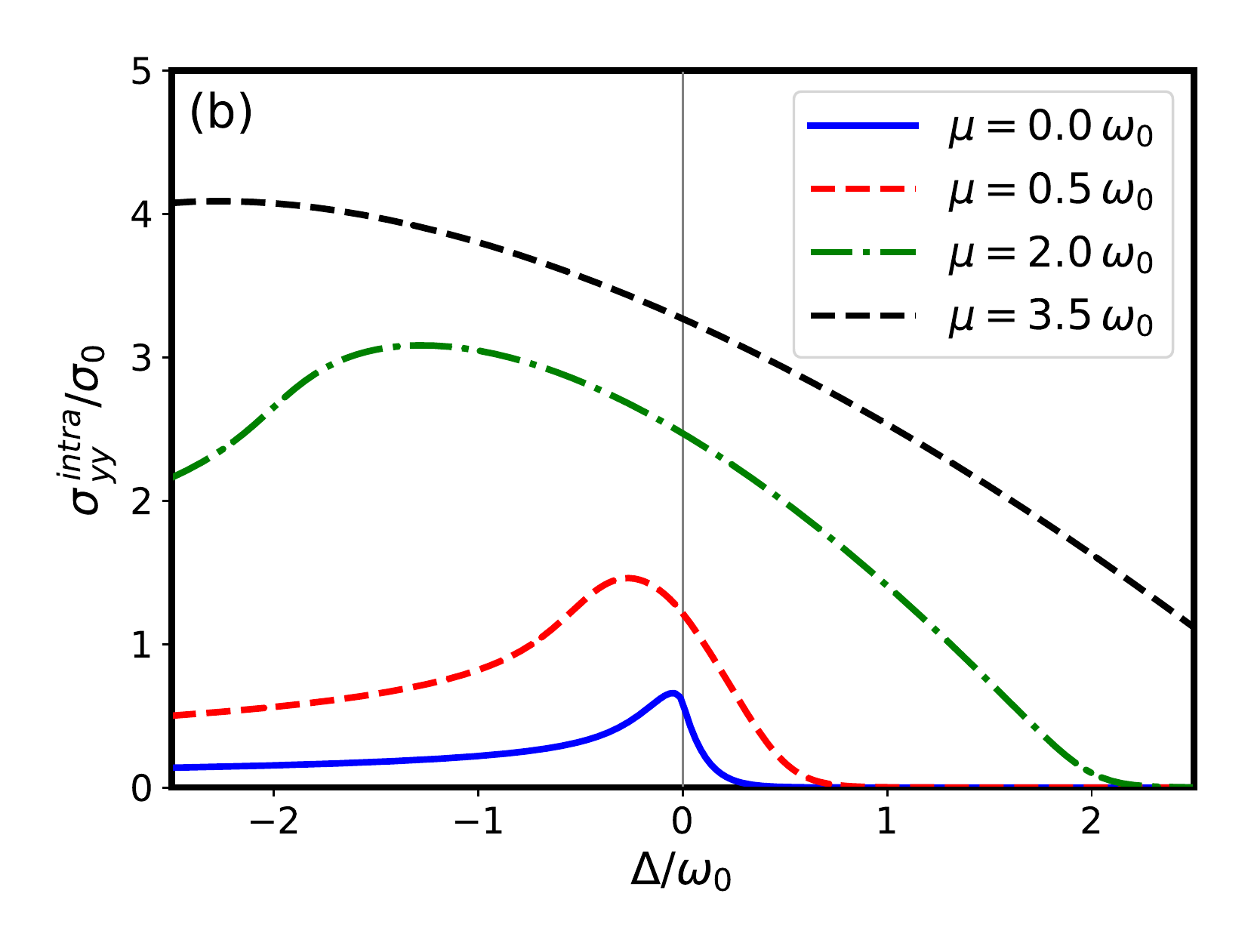}
	\caption{Real part of xx (a) and yy (b) intraband dc conductivities as functions of the gap $\Delta$ for different values of chemical potential.
The temperature is equal to $T=0.1\, \omega_0$ in both cases with $\omega_0=v^2/a$. The pronounced peak at $\mu=0$ in panel (b) manifests the possibility of dc transport through the charge-neutrality point.}
	\label{fig:sigma-semi-DC-Delta-dep}
\end{figure*}
For the longitudinal conductivity along the y-direction the technical details of calculation are very similar to the xx-case. They are presented in Appendix \ref{appendix:conductivity-semi-Dirac-xx-v2}. The results for interband ac optical conductivity are:
	\begin{align}\label{eq:Re-sigma-semi-yy-AC}
		&\operatorname{Re} \sigma_{yy}^{inter}(\omega)=\sgn\omega\frac{e^{2}}{ 2\pi \hbar}  \frac{v}{4\sqrt{2 |\omega| a}}
\left[f\left(-\frac{\omega}{2}\right)-f\left(\frac{\omega}{2} \right)\right]\times\nn
		&\times\begin{cases}
			\begin{array}{c}
			2\Theta(|\Delta|-|\omega/2|)I_{4}^{y y}(2\Delta/ |\omega|)+\\
			+2\Theta(|\omega/2|-|\Delta|)I_{2}^{y y}(2\Delta/|\omega|)
			\end{array},&\Delta<0,\\
			& \\
			 \frac{\Gamma^{2}\left(\frac{1}{4}\right)}{3 \sqrt{2\pi}},&\Delta=0,\\
			& \\
			2\Theta(|\omega/2|-\Delta)
			I_{2}^{yy}(2\Delta/|\omega|) ,&\Delta>0.
		\end{cases}
	\end{align}
They are presented in Fig.\ref{fig:sigma-semi-AC-w-dep} in lower panels for all three different
cases of $\Delta$. As is seen in the lower panel in Fig.\ref{fig:sigma-semi-AC-w-dep}(c), the optical conductivity in the y-direction diverges at the point $\omega=-2\Delta$ for $\Delta<0$. This divergence was also observed in numerical calculations in Refs.\cite{Carbotte2019PRB,Mawrie2019}. Using our exact expressions, we can derive asymptotic expansions in the integrals $I_{2}^{yy}(2\Delta/|\omega|)$ and $I_{4}^{yy}(2\Delta/|\omega|)$ at $\omega=2|\Delta|$ for negative $\Delta$. Expanding the integrals near this point up to leading order, we find:
\begin{align}
	&I_2^{yy}(2\Delta/|\omega|)_{\omega\to 2|\Delta|_{+}}\approx \frac{1}{\sqrt{2}}\log\frac{2|\Delta|}{\omega-2|\Delta|}+\text{const},\\
	&I_4^{yy}(2\Delta/|\omega|)_{\omega\to 2|\Delta|_{-}}\approx \frac{1}{\sqrt{2}}\log\frac{2|\Delta|}{|2\Delta|-\omega}+\text{const}.
\end{align}
The logarithmic singularity has the same amplitudes from both sides. In Ref.\cite{Carbotte2019PRB} this singularity was related to the joint
density of states for initial and final states involved in an optical transition, hence the van Hove singularity appears at $\omega=2|\Delta|$, while the
density of states itself has a van Hove logarithmic singularity at $\omega=|\Delta|$.   The density of states
for the considered system was derived in Ref.\cite{Montambaux2009EPJ}, it is expressed also in terms of complete elliptic integrals of the first and second kind.

We also present the asymptotes for the case $\Delta<0$ at small and large $\omega$:
\bea
\operatorname{Re} \sigma_{y y}^{inter}(\omega)\simeq \frac{e^{2}}{2\pi\hbar}\left\{\begin{array}{cc}\frac{v}{\sqrt{|\Delta|a}}
\frac{\pi\omega}{32T\cosh^2\frac{\mu}{2T}},
\quad \omega\to0,\\
\frac{v}{\sqrt{\omega a}}\frac{\Gamma^2\left(\frac{1}{4}\right)}{24\sqrt{\pi}},\quad\omega\to\infty.\end{array}\right.
\eea

For intraband dc optical conductivity we find
	\begin{align}\label{eq:Re-sigma-semi-yy-DC}
		&\operatorname{Re} \sigma_{yy}^{intra}(\omega)=\delta(\omega)\frac{e^2}{16\pi \hbar T}\int_{-\infty}^{\infty}  \frac{d E}{\cosh ^{2}
\left(\frac{E-\mu}{2 T}\right)} \frac{v\sqrt{|E|}}{\sqrt{a}}\times\nn
		&\times\begin{cases}
			\begin{array}{c}
			2\Theta(|\Delta|-|E|)I_{3}^{y y}(\Delta/|E|)+\\
			+2\Theta(|E|-|\Delta|)I_{1}^{y y}(\Delta/|E|)
			\end{array}, &\Delta<0,\\
			& \\
			 \frac{\sqrt{2}\Gamma^{2}\left(\frac{1}{4}\right)}{3 \sqrt{\pi}},&\Delta=0,\\
			& \\
			2\Theta(|E|-\Delta)
			I_{1}^{yy}(\Delta/|E|)
			,&\Delta>0.
		\end{cases}
	\end{align}
 Interband and intraband conductivities were studied recently in Ref.\cite{Carbotte2019} at zero temperature, the authors have obtained also asymptotic expressions at small and large frequencies. We checked that their asymptotics follow straightforwardly from our analytical results for $T=0$ while at finite temperature we
 get different dependence for $\operatorname{Re} \sigma_{yy}^{inter}(\omega)$ when $\omega$ goes to zero.

Finally, in Fig.\ref{fig:sigma-semi-DC-Delta-dep} we plot intraband parts as functions of the gap $\Delta$ for different values of chemical potential. The interesting feature presented in Fig.\ref{fig:sigma-semi-DC-Delta-dep}(b) is the appearance of a small peak near $\Delta=0$ on the negative side at small chemical potentials.
This peak can be related to the crossing of saddle point level with chemical potential. At zero chemical potential this peak appears only at small $\Delta$ values and attain maximum for $\Delta \approx 0$, which shows that temperature-broadened van Hove singularities intersect with the Fermi level and allow transport even at zero frequency.
Such signature can be used as a manifestation of the regime that is close to topological transition with $\Delta$ in dc transport measurements.

	\section{Optical conductivity of gapped dice model}
	
	\subsection{Solution of the Heisenberg equations for the quasiparticle in dice model}
	\label{sec:dice-Heisenberg}
\begin{figure*}	
		$(a)$
		\includegraphics[scale=0.35]{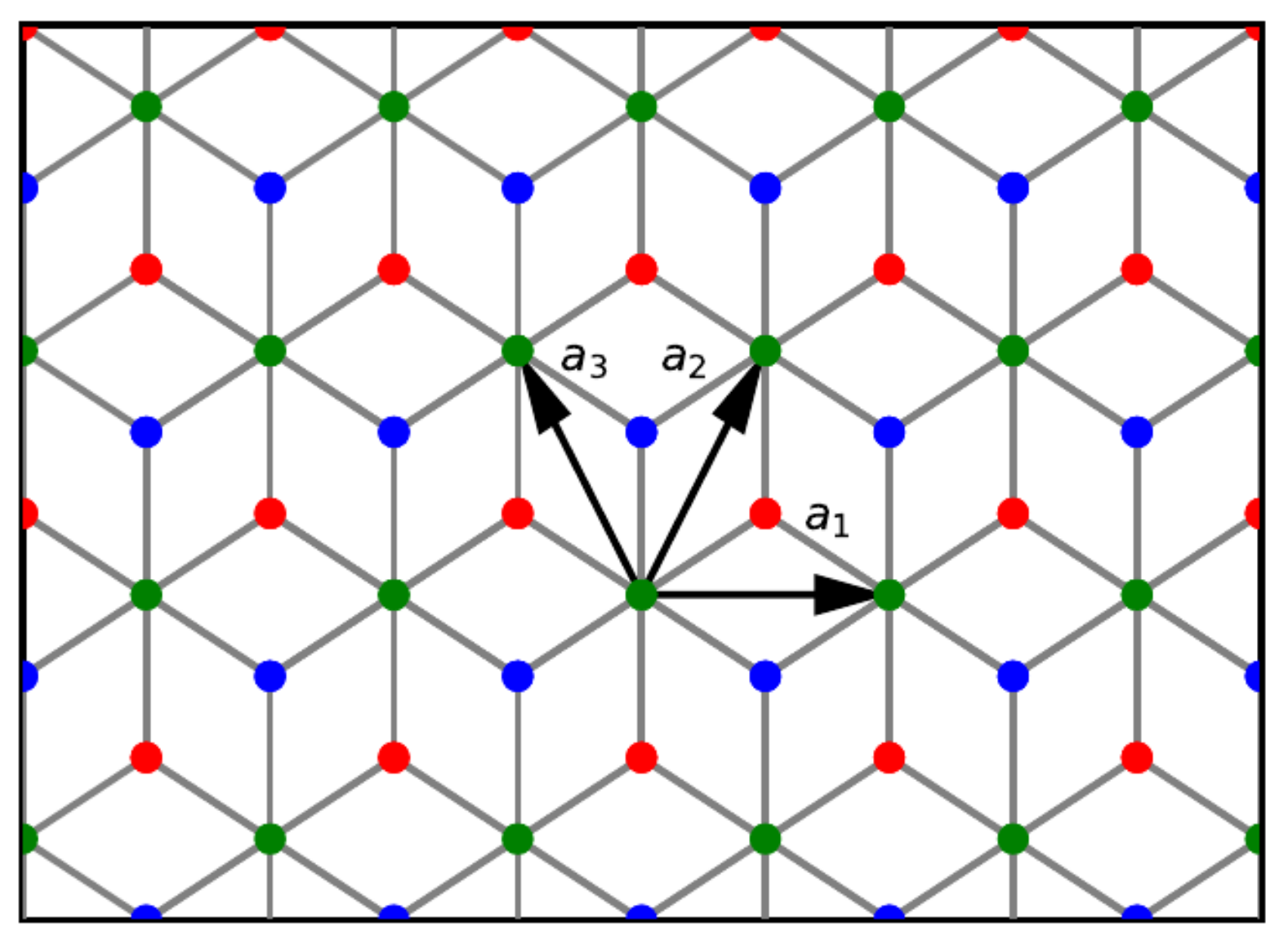}\quad
		$(b)$
		\includegraphics[scale=0.38]{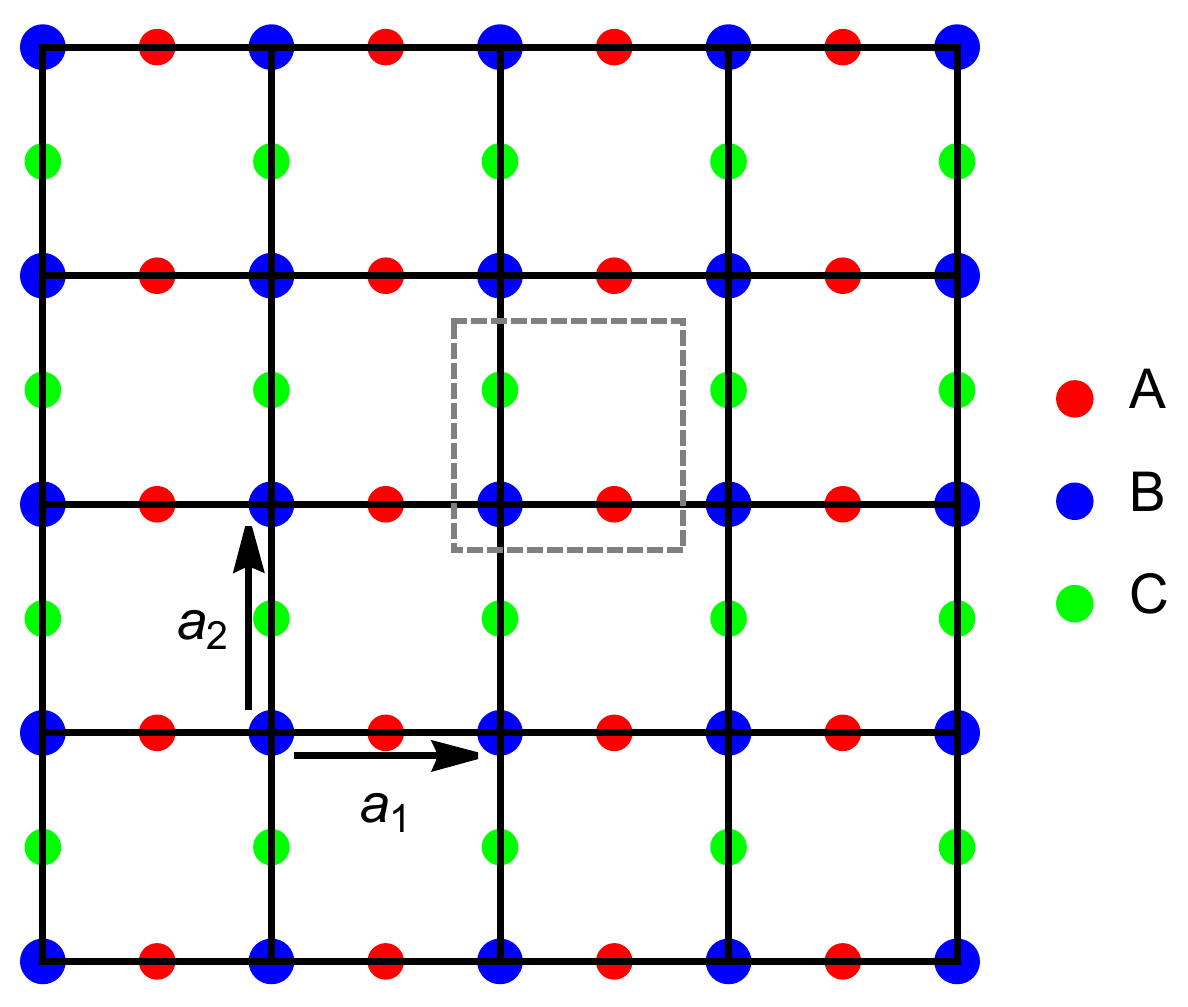}\quad
		$(c)$
		\includegraphics[scale=0.35]{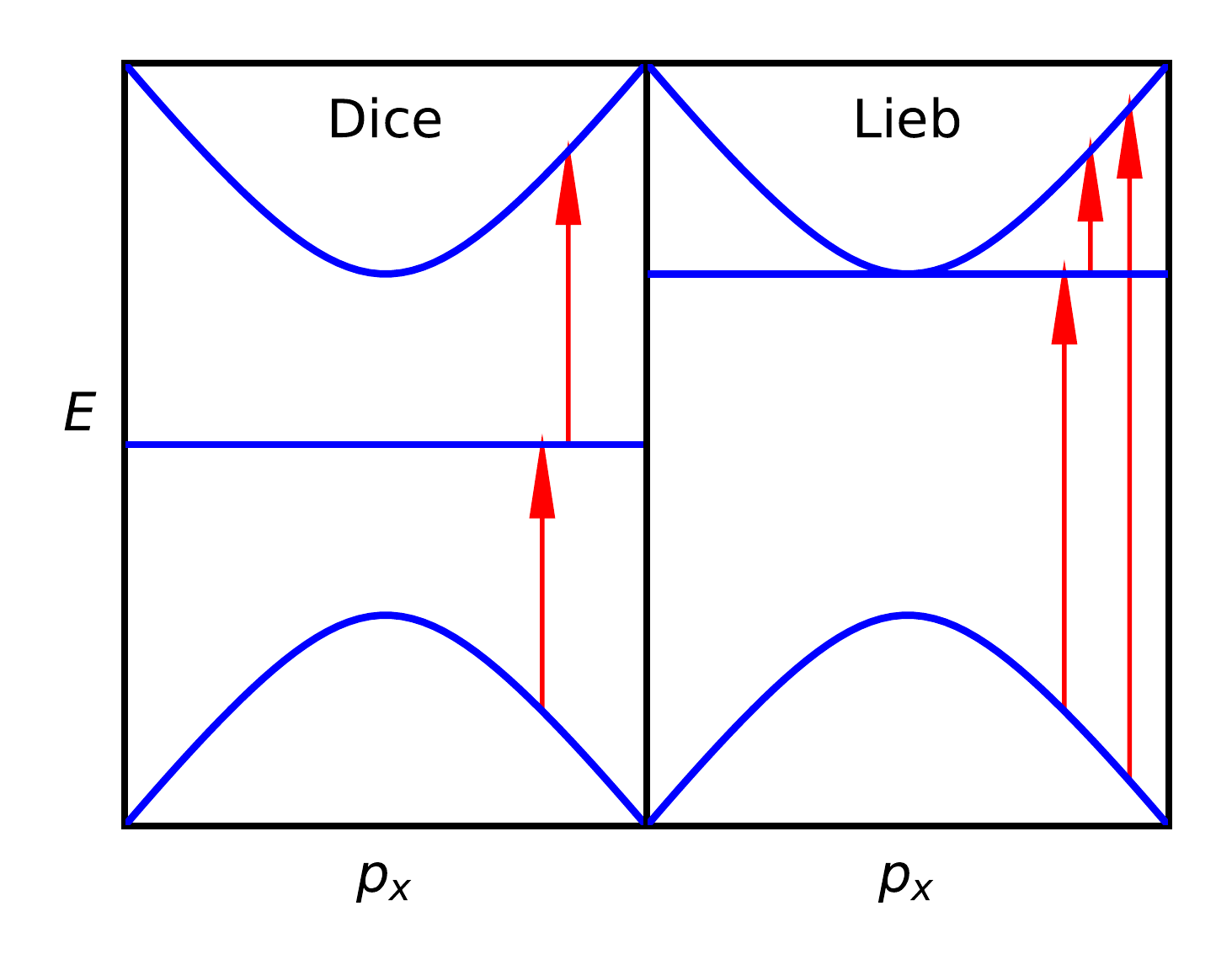}
		\caption{(a) A schematic plot of the lattice of the dice model. The red points display the $A$ sublattice atoms, the blue points describe the $B$ sublattice,
			and the green points define the $C$ sublattice. The vectors $\vec{a}_1=(1,\,0)a$ and $\vec{a}_2=(1/2,\,\sqrt{3}/2)a$ are the basis vectors of triangular sublattices. (b) The Lieb lattice with the corresponding sublattices, basis vectors and elementary cell. (c) Possible interband transitions which contribute to optical conductivity and define frequency thresholds.}
		\label{fig:dice}
	\end{figure*}

	The  $\mathcal{T}_{3}$ (dice) lattice is schematically shown in Fig.\ref{fig:dice}a. The corresponding tight-binding Hamiltonian is expressed through the function $f_{\mathbf{k}}=-\sqrt{2}t(1+e^{-i\mathbf{k}\mathbf{a}_2}+e^{-i\mathbf{k}\mathbf{a}_3})$ with equal hoppings $t$ between
	atoms $C$ (green hubs) and $A,B$ (red, blue rim sites) \cite{Vidal1998,Bercioux2009} and the corresponding energy spectrum is \cite{Raoux2014}
	\begin{align}
	&\epsilon_{0}=0,\nn
	&\epsilon_{\pm}=\pm\sqrt{2}t\bigg[3+2(\cos(\vec{a}_1\vec{k})+\cos(\vec{a}_2\vec{k})+\cos(\vec{a}_3\vec{k}))\bigg]^{1/2},
		\label{dice-model-energy-spectrum}
	\end{align}
	where $\vec{a}_1=(1,\,0)a$ and $\vec{a}_2=(1/2,\,\sqrt{3}/2)a$ are the basis vectors of the triangle sublattices and $\vec{a}_3=\vec{a}_2 - \vec{a}_1$ with the lattice constant denoted by $a$.
	
	There are two values of momentum where $f_{\vec{k}}=0$ and all three bands meet. They are situated at the corners of the hexagonal
	Brillouin zone
	\begin{align}
		K=\frac{2\pi}{a}\left(\frac{1}{3},\,\frac{1}{\sqrt{3}}\right),\quad K'=\frac{2\pi}{a}\left(-\frac{1}{3},\,\frac{1}{\sqrt{3}}\right).
	\end{align}
	For momenta near the $K$ and $K^{\prime}$ points, the function $f_{\vec{k}}$ is linear in $\vec{p}=\vec{k}-\xi\vec{K}$, i.e.,
	$f_{\vec{k}}=v_F(\xi p_x-ip_y)$, $v_F=\sqrt{3}ta/2$ is the Fermi velocity, and $\xi=\pm$ is the valley index. In addition, we set $\hbar=1$ for convenience.
	The low-energy Hamiltonian near K(K') $\xi=\pm 1$ three-band-touching point  reads:
	\begin{align}
		H_{dice}=v_F(p_x S_x+\xi p_y S_y+p_z S_z),
	\end{align}
	with a constant gap $v_Fp_z$ and pseudospin-1 matrices $S_i$ are
	\begin{align}
		&S_{x}=\frac{1}{\sqrt{2}}\left(\begin{array}{lll}
			0 & 1 & 0 \\
			1 & 0 & 1 \\
			0 & 1 & 0
		\end{array}\right),\,\,S_{y}=\frac{1}{\sqrt{2}}\left(\begin{array}{ccc}
			0 & -i & 0 \\
			i & 0 & -i \\
			0 & i & 0
		\end{array}\right),\nn
		&S^{z}=\left(\begin{array}{rrr}
			1 & 0 & 0 \\
			0 & 0 & 0 \\
			0 & 0 & -1
		\end{array}\right).
	\end{align}
	These matrices form a closed algebra with respect to commutator operation: $[S_i,S_j]=i\epsilon_{ijk}S_k$.
	
	The $S_z$-type term in the Hamiltonian $H_{dice}$ describes the spectral gap, which can be opened by adding on-site potential on $A$ and $B$ sites \cite{Gorbar2019}, in the Haldane model \cite{Dey2020} or dynamically generated in special cases of electron-electron interactions \cite{Gorbar2021a} and in the Floquet setup under circularly polarized radiation \cite{Dey2019gap,Iurov2019}.
	
	Let us perform analysis for K ($\xi=1$) valley, and then account for K' valley with proper sign changes. The Heisenberg equations for the coordinate and momentum operators in this case take the form:
	\begin{align}
		&\vec{v}(t)=\frac{d \vec{x}}{d t}=-i [ x(t),\,H_{dice}]=v_F\vec{S}(t),\\
		&\frac{d \vec{p}}{d t}=- i [p(t), \,H_{dice}]=0.
	\end{align}
Again, using the solution of the second equation, that states $p(t)=p(0)$, we arrive at the following Heisenberg equation for matrices $S_i$:
	\begin{align}
		&\frac{d S_i(t)}{d t}=-i[S_i(t),H_{dice}]=i P_{ij}S_j(t),
	\end{align}
	with
\bea
 P_{i j}=i  v_{F} \epsilon_{i j k} p_{k}= iv_{F}\left(\begin{array}{ccc}
			0 & p_z & -p_{y} \\
			-p_z & 0 & p_{x} \\
			p_{y} & -p_{x} & 0
		\end{array}\right).
\eea
	The solution of this equation has the form
	\begin{equation}\label{eq:S-matrix-solution}
		S_{i}(t)=\left(e^{i P t}\right)_{i j} S_{j}(0),
	\end{equation}
	where the matrix exponential is
	\begin{widetext}
	\begin{align}\label{eq:S-matrix-solution-matrix-exp}
		\left(e^{i P t}\right)_{i j}=\left(
		\begin{array}{ccc}
			\frac{\left(p_y^2+p_z^2\right) \cos \left(p t v_F\right)+p_x^2}{p^2} & \frac{p_x p_y \left(1-\cos \left(p t v_F\right)\right)-p p_z
\sin \left(p t v_F\right)}{p^2} & \frac{p_x p_z(1-\cos \left(p t v_F\right))+p p_y \sin \left(p t
				v_F\right)}{p^2} \\
			\frac{p_x p_y(1-\cos \left(p t v_F\right))+p p_z \sin \left(p t v_F\right)}{p^2} & \frac{\left(p_x^2+p_z^2\right) \cos \left(p t v_F\right)+p_y^2}{p^2} & \frac{p_y p_z(1-\cos \left(p t
				v_F\right))-p p_x \sin \left(p t v_F\right)}{p^2} \\
			\frac{p_x p_z(1-\cos \left(p t v_F\right))-p p_y \sin \left(p t v_F\right)}{p^2} & \frac{p p_x \sin \left(p t v_F\right)-p_y p_z \cos \left(p t v_F\right)+p_y p_z}{p^2} & \frac{\left(p_x^2+p_y^2\right) \cos \left(p t
				v_F\right)+p_z^2}{p^2} \\
		\end{array}
		\right).
	\end{align}
\end{widetext}
	Here we used the notation $p=\sqrt{p_x^2+p_y^2+p_z^2}$. The eigenvalues of the matrix $P$ are $\pm v_F p$, $0$. The matrix exponential greatly simplifies for the gapless case with $p_z=0$ (compare with Eq.\eqref{eq:solution-matrix-semi-Dirac}):
	\begin{align}
		&\left(e^{i P t}\right)_{i j}(p_z=0)=\nn
		&=\left(
		\begin{array}{ccc}
			\frac{p_y^2 \cos \left(p t v_F\right)+p_x^2}{p^2} & \frac{p_x p_y \left(1-\cos \left(p t v_F\right)\right)}{p^2} & \frac{p_y \sin \left(p t v_F\right)}{p} \\
			\frac{p_x p_y \left(1-\cos \left(p t v_F\right)\right)}{p^2} & \frac{p_x^2 \cos \left(p t v_F\right)+p_y^2}{p^2} & -\frac{p_x \sin \left(p t v_F\right)}{p} \\
			-\frac{p_y \sin \left(p t v_F\right)}{p} & \frac{p_x \sin \left(p t v_F\right)}{p} & \cos \left(p t v_F\right) \\
		\end{array}
		\right).
	\end{align}
	Thus, from the solutions \eqref{eq:S-matrix-solution} and \eqref{eq:S-matrix-solution-matrix-exp} we find the time-dependent velocity operators:
	\begin{align}
		v_{x}(t)=&v_F\left(\frac{\left(p_{y}^{2}+p_{z}^{2}\right) \cos \left(p t v_{F}\right)+p_{x}^{2}}{p^{2}}S_x+\right.\nn
		&\left.+\frac{p_{x} p_{y}(1- \cos \left(p t v_{F}\right))-p p_{z} \sin \left(p t v_{F}\right)}{p^{2}}S_y+\right.\nn
		&\left.+\frac{p_{x} p_{z}(1- \cos \left(p t v_{F}\right))+p p_{y} \sin \left(p t v_{F}\right)}{p^{2}}S_z\right),
	\end{align}
\begin{align}
		v_{y}(t)=&v_F\left(\frac{p_{x} p_{y}(1- \cos \left(p t v_{F}\right))+p p_{z} \sin \left(p t v_{F}\right)}{p^{2}}S_x+\right.\nn
		&\left.+\frac{\left(p_{x}^{2}+p_{z}^{2}\right) \cos \left(p t v_{F}\right)+p_{y}^{2}}{p^{2}}S_y+\right.\nn
		&\left.+\frac{p_{y} p_{z}(1- \cos \left(p t v_{F}\right))-p p_{x} \sin \left(p t v_{F}\right)}{p^{2}}S_z\right).
	\end{align}
	Below we insert these results into Eqs.\eqref{eq:Re-sigma-mu-nu} and \eqref{eq:Im-sigma-mu-nu}
	to evaluate the longitudinal and Hall conductivities. Again, we see that the velocities $v_i(t)$ contain zitterbewegung
terms which stem from the oscillating  terms.
	
	\subsection{Longitudinal and Hall conductivities in massive dice model}
	\label{sec:conductivity}
	Substituting the obtained velocities into Eqs.\eqref{eq:fixed-E-average},\eqref{eq:trace-Fourier-sE-general} and performing Fourier transform over pairs of $(s,E)$ and $(t,\omega)$ variables, we find
\begin{align}
&\mathcal{F}_{t,s}\Tr[e^{-iHs}v_x(t)v_x(0)]=\nn
&\pi v_F^2\delta (E) \left(\frac{p^2+p_z^2}{2p^2}\right)\left(\delta \left(\omega -p v_F\right)+\delta\left(\omega +p v_F\right)\right)+\nn
&+\pi v_F^2\delta\left(E+p v_F\right) \left(\frac{p^2+ p_z^2 }{2 p^2}\delta \left(\omega -p v_F\right)+\frac{p^2-p_z^2}{p^2}\delta (\omega )\right)+\nn
&+\pi v_F^2\delta \left(E-p v_F\right)\left(\frac{p^2+ p_z^2}{2 p^2}\delta\left(\omega +p v_F\right)+\frac{p^2-p_z^2}{p^2}\delta (\omega )\right),\\
&\mathcal{F}_{t,s}\Tr[e^{-iHs}v_{[x,}(t)v_{y]}(0)]=\frac{v_F^2 p_z}{i p} \times\nn
&\bigg[\delta(\omega- p v_F )\delta(E+p v_F)-\delta(\omega+p v_F )\delta(E-p v_F)\bigg.\nn
&\bigg.-\delta(E)\delta(\omega+p v_F )-\delta(\omega-p v_F )\bigg].\,
\end{align}
where the double Fourier transform is defined as
\be
\mathcal{F}_{t,s}f(t,s)=\int_{-\infty}^\infty\frac{dt\,ds}{(2\pi)^2}e^{i\omega t+i E s}f(t,s)
\ee
Using the first expression in the general formula for longitudinal conductivity, we find:
\begin{align}\label{eq:re-xx-dice}
&\operatorname{Re} \sigma_{x x}(\omega)=\frac{ e^{2} }{4\hbar} \left[\delta(\omega)\int_{-\infty}^{\infty} \frac{d E}{4T\cosh^{2}
\left(\frac{E-\mu}{2 T}\right)}\times\right.\nn
&\times\left.\frac{E^{2}-\Delta^2v_F^2}{|E|} \Theta\left(|E|-\Delta v_F\right)+\right.\nn
&+\left.\frac{ \omega^2+\Delta^2v_F^2}{2 \omega^2}\Theta\left(|\omega|-\Delta v_F\right)[f(-|\omega|)-f(|\omega|)]\right],
\end{align}
where we relabeled $p_z=\Delta>0$ and took into account the presence of two valleys that contribute equally. Note that the term proportional to $\Theta(|\omega|-\Delta v_F)$ defines the energy threshold after which the transitions from and to flat band become possible. However, no special threshold is present for transitions between the two dispersive bands, which means that only transitions through flat band are possible. This was already pointed out for the gapless dice model in Refs.\cite{Illes2015,Han2022PRB}.

\begin{figure*}
		\centering
		\includegraphics[scale=0.34]{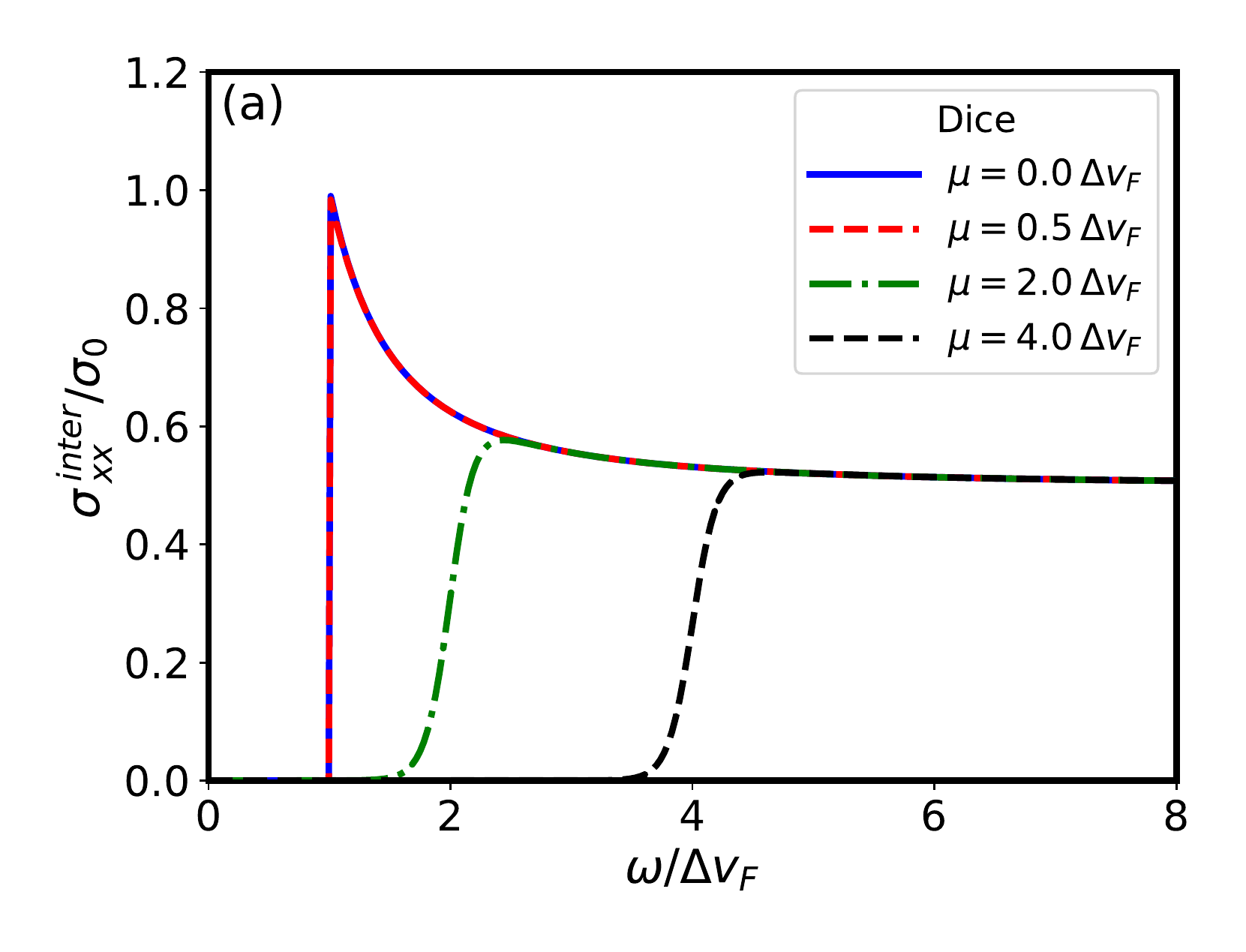}
		\includegraphics[scale=0.34]{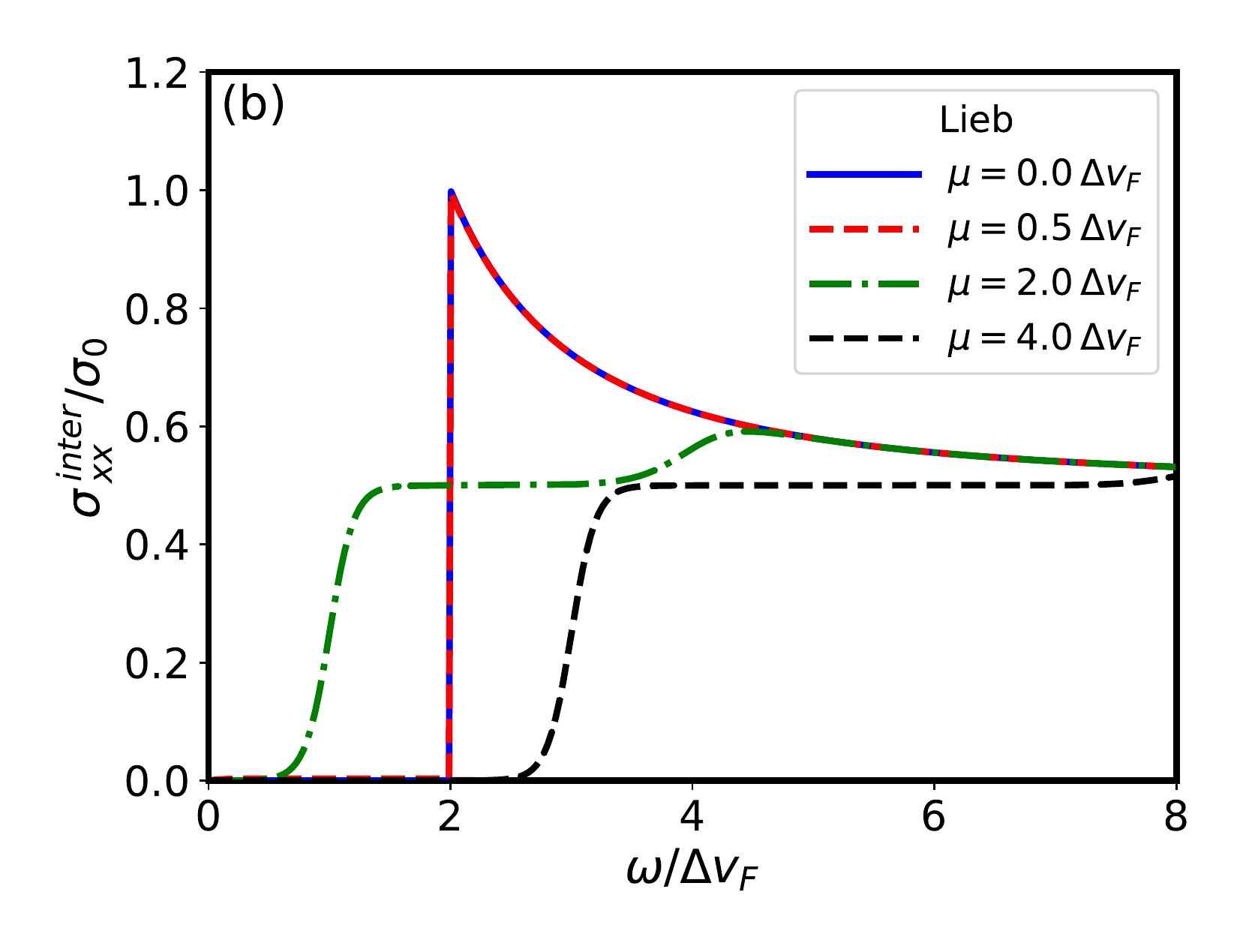}
		\includegraphics[scale=0.34]{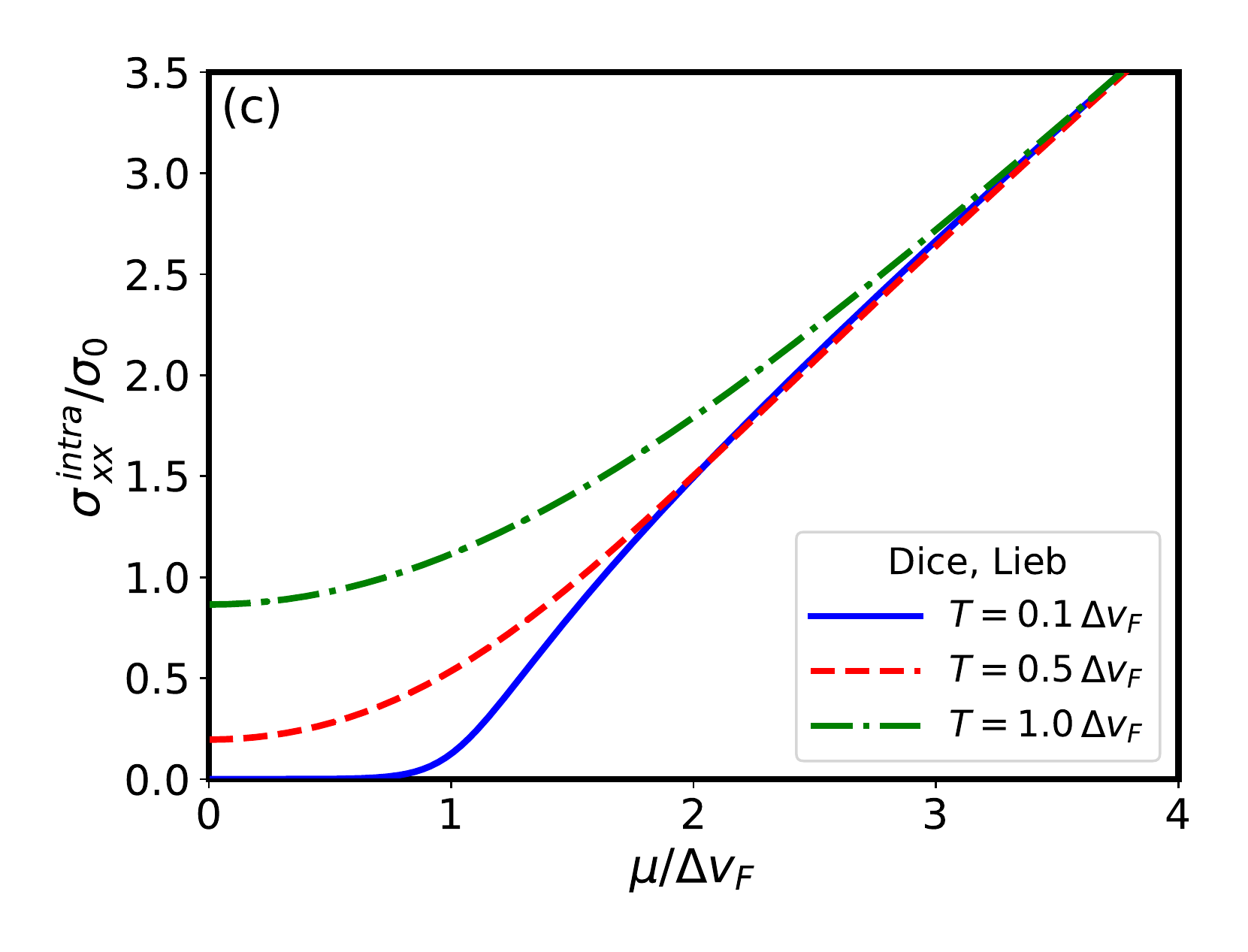}
		\caption{Panels (a) and (b): the real part of optical conductivity for gapped dice and Lieb lattices given by Eqs.\eqref{eq:re-xx-dice} and \eqref{eq:re-sigma-Lieb} at temperature $T=0.1 \Delta v_F$. Panel (c): the real part of intraband dc conductivity which is the same for both lattices (for dice lattice
in a single valley).}
		\label{fig:conductivities-dice}
	\end{figure*}

Similarly, for the imaginary part of the Hall conductivity in one valley we find
	\begin{align}\label{eq:Hall-conductivity-dice}
		\operatorname{Im} \sigma_{[x, y]}(\omega)=\frac{e^{2}p_z v_F}{4\hbar \omega} \Theta\left(|\omega|-v_{F} |p_{z}|\right)[f(|\omega|)-f(-|\omega|)].
	\end{align}
Note that the Hall conductivity is proportional to the gap parameter $p_z$ and the sum over two valleys with different signs of $p_z$ will lead
to the zero total Hall conductivity.
This is because the system is T-invariant, and the operation of T-invariance interchanges K and K' valleys \cite{Gorbar2019}.
These conductivities are shown in Fig.\ref{fig:conductivities-dice} for different values of chemical potential and temperature.

Using the Kramers-Kronig relations, one can evaluate the real part of the Hall conductivity, see Eq.\eqref{eq:integral-Hall-Re-dice}. At zero temperature we find the following expression:
\begin{align}
	\operatorname{Re} \sigma_{x y}(\omega)=-\frac{e^{2} v_F p_z}{4 \pi \hbar \omega} \log \left|\frac{\max (|\mu|,v_F |p_z|)+\omega}
{ \max (|\mu|,v_F |p_z|)-\omega}\right|.
\label{Hall-dice-T0}
\end{align}
At the energy $\omega=\mbox{max}(|\mu|,v_F|p_z|)$, there is a logarithmic divergence in the Hall conductivity. For large energies, $\omega\to\infty$, this expression approaches zero as $\sim 1/\omega^2$.
	This expression is very similar to those obtained in graphene-like systems (see, for example, \cite{Li2012,Tabert2013}). The dc limit $\omega\to0$
leads to the quantized Hall conductivity ${\rm Re}\,\sigma_{xy}=-e^2{\rm sign}(p_z)/h$ for $|\mu|\le v_F|p_z|$ in the absence of a magnetic field \cite{Sinitsyn2006PRL}.
	
	\section{Optical conductivity of the Lieb model}
	\label{sec:Lieb-optical}
	In this section we evaluate the optical conductivity of the gapped Lieb model \cite{Shen2010} using the method presented above. The main complication arises in solving Heisenberg equations for matrices: due to commutation relations the whole set of the Gell-Mann matrices enters the calculation. Below we show how one can still perform calculation and arrive at relatively simple expression for the conductivity.
	 We start with description of the main properties of the Lieb lattice and corresponding low-energy model.

	\subsection{Lieb lattice and low-energy model}
The Lieb lattice is schematically shown in Fig.\ref{fig:dice}b. It consists of three square sublattices, with atoms placed in the corners and in the middle of each side of big squares forming a line-centered-square lattice. The tight-binding Hamiltonian, described in Ref.\cite{Shen2010}, reduces to the following low-energy model near the center of BZ $k_{x,y}=\frac{\pi}{a}+q_{x,y}$:
	\begin{align}
	H_{Lieb}=\left(
	\begin{array}{ccc}
		\Delta  v_F & v_F q_x & 0 \\
		v_F q_x & -\Delta  v_F & v_F q_y \\
		0 & v_F q_y & \Delta  v_F \\
	\end{array}
	\right),
\end{align}
where the site energies are set as $\epsilon_B=\epsilon_C=-\epsilon_A=\Delta v_F$. In terms of the Gell-Mann $\lambda$-matrices the Hamiltonian takes the form
\be
H_{Lieb}=v_F\left[\lambda_1q_x+\lambda_6 q_y+\Delta\left(\frac{\lambda_0}{3}+\lambda_3-\frac{\lambda_8}{\sqrt{3}}\right)\right].
\ee
Here $\lambda_0$ is the $3\times3$ unit matrix.
The energy dispersions defined by this Hamiltonian are given by three bands, one is flat band and the other two are dispersive bands (see Fig.\ref{fig:dice}c):
\begin{align}
	\epsilon_{0}=\Delta v_F,\,\,\epsilon_{\pm}=\pm v_F \sqrt{\Delta ^2+q_x^2+q_y^2}.
\end{align}
Let us check the T-invariance of this Hamiltonian. The operator $T$ should contain complex conjugation, the change of the sign of both momenta and contain the proper matrix transformation in sublattice space:
\begin{align}
	\hat{T} H(\boldsymbol{q}) \hat{T}^{-1}=H(-\boldsymbol{q}),\quad \hat{T}=F \hat{K}.
\end{align}
In the absence of the gap the matrix $F$ has the form
\begin{align}
	F=\begin{pmatrix}
		1 & 0 & 0\\
		0 & -1 & 0\\
		0 & 0 & 1
	\end{pmatrix}.
\end{align}
Thus we conclude that the gap presented in Ref.\cite{Shen2010} does not break T-invariance. Consequently, the Hall conductivity is zero in this model
in the absence of a magnetic field.
	
	\subsection{Solution of the Heisenberg equations}

	The Heisenberg equations for the coordinate and momentum operators are very similar to that obtained in previous sections: velocities evolve with time as the corresponding matrices in the Hamiltonian near $q_x$ and $q_y$, and the momenta do not evolve at all. The nontrivial part comes from the equation that describes the evolution  of matrices.
	The system of equations for the Gell-Mann matrices has the form:
\begin{align}
		\frac{d\lambda_i(t)}{dt}=-i[\lambda_i(t),H_{Lieb}]=v_FA_{ij}\lambda_j(t),
\label{eq:lambda-matrices}
\end{align}
where we used the commutation relations $[\lambda_i,\lambda_k]=2i f_{ikj}\lambda_j$ with $f_{ikj}$ being the structure constants of the $su(3)$ algebra, hence the matrix $A_{ij}$ has the form:
\begin{align} 	
	    A=\left(
		\begin{array}{cccccccc}
			0 & -2 \Delta  & 0 & 0 & q_y & 0 & 0 & 0 \\
			2 \Delta  & 0 & -2 q_x & -q_y & 0 & 0 & 0 & 0 \\
			0 & 2 q_x & 0 & 0 & 0 & 0 & -q_y & 0 \\
			0 & q_y & 0 & 0 & 0 & 0 & -q_x & 0 \\
			-q_y & 0 & 0 & 0 & 0 & q_x & 0 & 0 \\
			0 & 0 & 0 & 0 & -q_x & 0 & 2 \Delta  & 0 \\
			0 & 0 & q_y & q_x & 0 & -2 \Delta  & 0 & -\sqrt{3} q_y \\
			0 & 0 & 0 & 0 & 0 & 0 & \sqrt{3} q_y & 0 \\
		\end{array}
		\right).
	\end{align}
	
	For the eigenvalues of the matrix $v_F A_{ij}$ we find:
	\begin{align}
		&a_{1,2}=0,\quad a_{3,4}=\pm 2 i p v_F
		\nn
		&a_{5,6}=\pm i v_F (\Delta +p),\quad a_{7,8}=\pm i v_F (p-\Delta ),
	\end{align}
where we defined $p=\sqrt{q_x^2+q_y^2+\Delta^2}$. The initial conditions for velocities are $v_x(0)=v_F \lambda_1,\quad v_y(0)=v_F \lambda_6$.
	After calculation of the matrix exponent $\exp[A t]$, we find velocities at time $t$ by taking the corresponding rows in resulting matrix - the first for $v_x$ and the sixth for $v_y$. The solutions for $v_x$ and $v_y$ are defined as vectors in the Gell-Mann basis - see Eqs.\eqref{eq:velocity-x-Gell-Mann} and \eqref{eq:velocity-y-Gell-Mann} in Appendix \ref{appendix:Lieb-conductivity-details}. The identity matrix is not present because it does not evolve with time and the coefficient before this matrix is zero.
	Next we evaluate the conductivity using the obtained solutions $v_{x,y}(t)$ and previously established method.
	
	\subsection{Optical conductivity}
	Performing trace evaluation and using the double-Fourier transform, we arrive at the following final answer for the optical conductivity of the Lieb lattice in the x-direction (see Appendix \ref{appendix:Lieb-conductivity-details}):
		\begin{align}\label{eq:re-sigma-Lieb}
&\Re\sigma_{xx}(\omega)=\frac{e^2 }{4\hbar}\left[\delta(\omega)\int_{-\infty}^{\infty} \frac{d E}{4T\cosh^{2}
	\left(\frac{E-\mu}{2 T}\right)}\times\right.\nn
&\times\left.\frac{E^{2}-\Delta^2v_F^2}{|E|} \Theta\left(|E|-\Delta v_F\right)+\right.\nn
&\left.+\Theta(|\omega|-2\Delta v_F)\left[\frac{2\Delta^2 v_F^2}{\omega^2}\left(f\left(-\frac{|\omega|}{2}\right)
-f\left(\frac{|\omega|}{2}\right)\right)+\right.\right.\nn
&\left.\left.+\frac{f(\Delta v_F-|\omega|)-f(\Delta v_F)}{2}\right]+\frac{f(\Delta v_F)-f(\Delta v_F+ |\omega|)}{2} \right].
\end{align}
Performing calculation for the y-direction we find the same answer.
	
	The physical meaning of the terms in Eq.\eqref{eq:re-sigma-Lieb} is the following: the first term corresponds to intraband dc conductivity, the second term describes interband transitions through the gap - that is why the threshold is $2\Delta v_f$, and the last term corresponds to transitions between flat and upper dispersive band.
	This conductivity is presented in Fig.\ref{fig:conductivities-dice}.
	
	The interesting difference compared to the dice model conductivity \eqref{eq:re-xx-dice} is the presence of both dispersive-to-dispersive band transitions and dispersive-to-flat band transitions in the interband ac part of optical conductivity (schematically shown in Fig.\ref{fig:dice}c).
	
	\section{Conclusions}
	In the present paper we further developed the approach of Refs.\cite{Katsnelson2006EPJB,Cserti2006PRB} for calculating longitudinal and Hall conductivities of systems with arbitrary pseudospin and dispersion law of quasiparticles. The conductivities are written through  quasiparticle velocity correlators at time $t$ for states of energy $E$ which also describe the phenomenon of zitterbewegung. For noninteracting systems the Heisenberg equations for velocities can be solved that allows one to significantly reduce the complexity of the conductivity calculation and obtain in some cases closed-form analytic expressions. 

	
	We applied this method to evaluate the optical conductivity of the semi-Dirac model, which is an example of low-energy theory with anisotropic spectrum. We obtained exact expressions which allowed us to identify the signatures of topological phase transition with gap closing and merging Dirac points. The previously unobserved result is the peak in the intraband dc conductivity along the y-direction at zero chemical potential when the two Dirac cones nearly merge with each other. Physically, one would expect that this is related to the intersection of  broadened van Hove singularities with the Fermi level. Such an intersection leads to the appearance of a number of propagating states carrying a nonzero current. At low temperatures, nonzero transport through the charge-neutrality point can be used as a manifestation of topological transition regime.
	
	In addition, we analyzed two gapped pseudospin-1 models that correspond to dice and Lieb lattices. The optical conductivities for the considered gap parameters were not studied previously. The key physical difference that we observed is the fact that in the gapped Lieb model all transitions between three bands (dispersive-to-flat, flat-to-dispersive and between two dispersive) contribute to the optical conductivity at large frequencies, while in dice lattice only transitions to- and from flat band play a role.

\acknowledgements
We are grateful to E.V. Gorbar and Y. Cheipesh for useful remarks. D.O.O. acknowledges the support from the Netherlands Organization for Scientific Research (NWO/OCW)
and from the European Research Council (ERC) under the European Union’s Horizon 2020 research and innovation
programme. V.P.G. acknowledges support by the National Academy of Sciences of Ukraine grant ``Effects of external fields and spatial inhomogeneities on the electronic properties of Dirac and superconducting materials'' (Grant No.  0122U002313).

	\newpage
	\appendix
	\onecolumngrid
	\section{Derivation of general conductivity expressions from Kubo formula}
	\label{appendix:A}

\subsection{Expression of the conductivity tensor through retarded correlation function}

It is well known that the conductivity (\ref{Kubo-formula}) can be written through the Fourier transform of the retarded correlation function
$\Pi_{\mu \nu}^{r}(t)=-i \theta(t)\left\langle\left[J_{\mu}(t)J_{\nu}(0)\right]\right\rangle$:
\begin{align}
	&\sigma_{\mu \nu}(\omega)=\frac{i K_{\mu \nu}(\omega+i \epsilon)}{\omega+i \epsilon}, \nn &K_{\mu \nu}(\omega+i \epsilon)=\frac{\langle\tau\rangle}{V} \delta_{\mu \nu}+\frac{\Pi_{\mu \nu}^{r}(\omega+i \epsilon)}{V}.
\label{tau-term}
\end{align}
The function $\Pi_{\mu \nu}^{r}(\omega)$ can be obtained by analytical continuation from its imaginary time expression $\left(\Pi_{\mu \nu}^{r}(\omega)=\right.$ $\left.\Pi_{\mu \nu}\left(i \omega_{m} \rightarrow \omega+i \epsilon\right)\right)$.
For noninteracting fermions, using the Matsubara diagram technique for evaluating $\tau$-ordered product of operators we get
\begin{equation}
	\Pi_{\mu \nu}\left(i \omega_{m}\right)=\frac{1}{\beta} \sum_{n=-\infty}^{\infty} \operatorname{Tr}\left[j_{\mu} \frac{1}{i \Omega_{n}-H_{0}} j_{\nu} \frac{1}{i \Omega_{n}-i \omega_{m}-H_{0}}\right].
\end{equation}
In the energy representation it takes the form
\begin{align}
\Pi_{\mu \nu}\left(i \omega_{m}\right)=\frac{1}{\beta} \sum_{\alpha, \beta} j_{\mu}^{\alpha \beta} j_{\nu}^{\beta \alpha} \sum_{n=-\infty}^{\infty} \frac{1}{\left(i \Omega_{n}-E_{\beta}\right)\left(i \Omega_{n}-i \omega_{m}-E_{\alpha}\right)}.
\end{align}
The summation over the Matsubara frequencies can be easily performed, thus we get
\begin{align}
	\Pi_{\mu \nu}\left(i \omega_{m}\right)=\sum_{\alpha, \beta} j_{\mu}^{\alpha \beta} j_{\nu}^{\beta \alpha} \frac{f\left(E_{\alpha}\right)-f\left(E_{\beta}\right)}{E_{\alpha}-E_{\beta}+i \omega_{m}},
\end{align}
where $f(E)$ is the Fermi-Dirac distribution function, $f(E)=1 /(\exp (\beta(E-\mu))+1)$. We now write
\begin{equation}
	J_{\mu}^{\alpha \beta} J_{\nu}^{\beta \alpha}=J_{\{\mu}^{\alpha \beta} J_{\nu\}}^{\beta \alpha}+J_{[\mu}^{\alpha \beta} J_{\nu]}^{\beta \alpha},
\end{equation}
where $J_{\{\mu} J_{\nu\}} \equiv\left(J_{\mu} J_{\nu}+J_{\nu} J_{\mu}\right) / 2$ and $J_{[\mu} J_{\nu]} \equiv\left(J_{\mu} J_{\nu}-J_{\nu} J_{\mu}\right) / 2$ denote symmetric and antisymmetric parts of the tensor $J_{\mu} J_{\nu}$, respectively. Using hermiticity of the current it is easy to show that the symmetric part $J_{\{\mu} J_{\nu\}}$ is a real quantity while the antisymmetric part $J_{[\mu} J_{\nu]}$ is the purely imaginary one. Therefore, after performing analytical continuation over frequency, we find the real symmetric part of $\sigma_{\mu \nu}$,
\begin{align}
\operatorname{Re} \sigma_{\{\mu, \nu\}}(\omega)=\frac{\pi e^{2}}{V \omega} \sum_{\alpha, \beta} v_{\{\mu}^{\alpha \beta}
v_{\nu\}}^{\beta \alpha}\left[f\left(E_{\alpha}\right)-f\left(E_{\beta}\right)\right] \delta\left(E_{\alpha}-E_{\beta}+ \omega\right),
\end{align}
where we used the relation $j_{\mu}=-e v_{\mu}$ between the current density and the velocity $(e>0)$. Accordingly, for the imaginary antisymmetric part of $\sigma_{\mu \nu}$ we have
\begin{align}
\operatorname{Im} \sigma_{[\mu, \nu]}(\omega)&=\frac{\pi e^{2}}{V \omega} \sum_{\alpha, \beta} \operatorname{Im}\left(v_{[\mu}^{\alpha \beta}
v_{\nu]}^{\beta \alpha}\right)\left[f\left(E_{\alpha}\right)-f\left(E_{\beta}\right)\right]\delta\left(E_{\alpha}-E_{\beta}+ \omega\right).
\end{align}
To restore remaining imaginary and real parts we can use the Kramers-Kr\"{o}nig relationships,
\begin{align}\label{eq:Kramers-Kronig}
	&\operatorname{Im} \sigma_{\{\mu, \nu\}}(\Omega)=-\frac{1}{\pi} \text { P.v. } \int_{-\infty}^{\infty} \frac{d \omega \operatorname{Re} \sigma_{\{\mu, \nu\}}(\omega)}{\omega-\Omega}, \nn
	&\operatorname{Re} \sigma_{[\mu, \nu]}(\Omega)=\frac{1}{\pi} \text { P.v. } \int_{-\infty}^{\infty} \frac{d \omega \operatorname{Im} \sigma_{[\mu, \nu]}(\omega)}{\omega-\Omega}.
\end{align}
Writing
\begin{align}
	\delta\left(E_{\alpha}-E_{\beta}+\omega\right)=\int_{-\infty}^{\infty} d E \delta\left(E-E_{\alpha}\right) \delta\left(E-E_{\beta}+ \omega\right)
\end{align}
we have for the symmetric part 
\begin{eqnarray}\label{eq:Re-sigma-munu-general-2}
	\operatorname{Re} \sigma_{\{\mu, \nu\}}(\omega) &=&\frac{\pi e^{2}}{V \omega} \sum_{\alpha, \beta} \int_{-\infty}^{\infty}
d E v_{\{\mu}^{\alpha \beta} v_{\nu\}}^{\beta \alpha} \delta\left(E-E_{\alpha}\right)
\delta\left(E-E_{\beta}+\omega\right)\left[f\left(E_{\alpha}\right)-f\left(E_{\beta}\right)\right] \nonumber\\
&=&\frac{\pi e^{2}}{V\omega} \int_{-\infty}^{\infty} d E [f(E-\omega)-f(E)]\operatorname{Tr}\left[v_{\{\mu} \delta(E-H) v_{\nu\}} \delta(E-H- \omega)\right].
\end{eqnarray}
In the last line we replaced the eigenvalues $E_{\alpha,\beta}$ by the Hamiltonian and sum over eigenstates by the trace over quantum numbers describing the system eigenstates.
Similarly, for the imaginary antisymmetric part we find:
\begin{align}\label{eq:Im-sigma-munu-general-2}
\operatorname{Im} \sigma_{[\mu, \nu]}(\omega)=\frac{\pi e^{2}}{V\omega} \int_{-\infty}^{\infty} d E [f(E-\omega)-f(E)]
\Im \operatorname{Tr}\left[v_{[\mu} \delta(E-H) v_{\nu]} \delta(E-H- \omega)\right].
\end{align}
Using the relation between traces and velocity correlators averaged at fixed energy (see Sec\ref{sec:trace}), we find the results presented in the main text, Eqs.\eqref{eq:Re-sigma-mu-nu} and \eqref{eq:Im-sigma-mu-nu}.

\subsection{Relation between trace and time-dependent velocity operators}
\label{sec:trace}
Let us consider the term $
	\operatorname{Tr}\left[v_{\mu} \delta(E-H) v_{\nu} \delta(E-H-\omega)\right]$
in the expressions \eqref{eq:Re-sigma-munu-general-2} and \eqref{eq:Im-sigma-munu-general-2} for interband ac conductivity. Also, $J_{\mu}(t)$ is the
actual current measured experimentally, the corresponding total current-density is obtained by differentiating
the Hamiltonian with respect to the vector potential,
\begin{align}
	J_{\mu}(\mathbf{r}, t)=-\frac{\delta H}{\delta\left(A_{\mu}(\mathbf{r}, t) / c\right)}.
\end{align}
Using the representation for the first delta function,
\begin{equation}
	\delta(E-H)=\frac{1}{2 \pi} \int_{-\infty}^{\infty} d t e^{i(E-H) t },
\end{equation}
and the cyclic property of a trace, then changing the variable of integration $E \rightarrow E+\omega$, we can write
\begin{equation}
	\operatorname{Tr}\left[v_{\mu} \delta(E-H) v_{\nu} \delta(E-H-\omega)\right]=\frac{1}{2 \pi } \int_{-\infty}^{\infty} d t e^{i \omega t} \operatorname{Tr}\left[\delta(E-H) v_{\mu}(t) v_{\nu}(0)\right].,
\end{equation}
Defining the microcanonical average of an operator $\hat{A}$ at given energy $E$,
\begin{equation}
	\langle\hat{A}\rangle_{E}=\frac{\operatorname{Tr}[\delta(E-\hat{H}) \hat{A}]}{\operatorname{Tr}[\delta(E-\hat{H})]},
\end{equation}
where $\operatorname{Tr}[\delta(E-\hat{H})]=\rho(E) V$ is the total density of states (DOS), we get the following expression for the symmetric ac conductivity through the correlator of velocities:
\begin{equation}\label{eq:Re-sigma-mu-nu-v2}
	\operatorname{Re} \sigma_{\{\mu, \nu\}}(\omega)=\frac{e^{2}}{2\omega} \int_{-\infty}^{\infty} d E \rho(E)\left[f(E)-f(E+ \omega)\right] \int_{-\infty}^{\infty} d t e^{i \omega t}\left\langle v_{\{\mu}(t) v_{\nu\}}(0)\right\rangle_{E}.
\end{equation}
It is easy to check the reality of the last expression using the relationship $\left\langle v_{\{\mu}(-t) v_{\nu\}}(0)\right\rangle_{E}^{*}=\left\langle v_{\{\mu}(t) v_{\nu\}}(0)\right\rangle_{E}$.

The expression \eqref{eq:Re-sigma-mu-nu} for $T=0$ is in accordance with Ref.\cite{Mayou2000PRL} for diagonal conductivity.
Similarly, for the imaginary antisymmetric part of conductivity we obtain
\begin{equation}
	\left.\operatorname{Im} \sigma_{[\mu, \nu]}(\omega)=\frac{e^{2}}{2 \omega} \operatorname{Im} \int_{-\infty}^{\infty} d E \rho(E)[f(E)-f(E+\omega))\right] \int_{-\infty}^{\infty} d t e^{i \omega t}\left\langle v_{[\mu}(t) v_{\nu]}(0)\right\rangle_{E}.
\end{equation}
To calculate $\operatorname{Im} \sigma_{\{\mu, \nu\}}(\omega)$ and $\operatorname{Re} \sigma_{[\mu, \nu]}(\omega)$ we use the Kramers-Kr\"{o}nig relation \eqref{eq:Kramers-Kronig}.

\section{Momentum integration in expressions for conductivity of the semi-Dirac model.}
\label{appendix:conductivity-semi-Dirac-xx-v2}
In this Appendix we discuss technical details regarding evaluation of longitudinal conductivity in the semi-Dirac model.
Following Ref.\cite{Gusynin2007conductivity}, one can express the diamagnetic term $\la\tau_{\mu\mu}\ra$ appearing in Eq.\eqref{Kubo-formula} as
\begin{align}
	\frac{\left\langle\tau_{\alpha \alpha}\right\rangle}{V}=e^{2} \int_{B Z} \frac{d^{2} p}{(2 \pi)^{2}} \frac{1}{2 \varepsilon(\mathbf{p})}\left[f(\epsilon_{+}(\mathbf{p}))-f(-\epsilon_{+}(\mathbf{p}))\right]\left(\Phi(\mathbf{p}) \frac{\partial^{2}}{\partial p_{\alpha}^{2}} \Phi^{*}(\mathbf{p})+\text { c.c. }\right),
\end{align}	
where $\Phi(\vec{p})$ is defined by model Hamiltonian \eqref{eq:semi-1} as
\begin{align}
	H_{semi}=\begin{pmatrix}
		0 & \Phi(\vec{p})\\
		\Phi^{*}(\vec{p}) &0
	\end{pmatrix},\quad \Phi(\vec{p})=\left(\Delta+a p_{x}^{2}\right)-i v p_{y}.
\end{align}
Thus, only the $\la\tau_{xx}\ra$ contribution is nonzero. After substituting the exact form of the dispersion and taking derivative of $\Phi(\vec{p})$, we find that the term $\la\tau_{xx}\ra$ is real:
 \begin{equation}
 	\frac{\left\langle\tau_{x x}\right\rangle}{V}=e^{2} \int \frac{d^{2} \mathbf{p}}{(2 \pi)^{2}} \frac{2 a\left(\Delta+a p_{x}^{2}\right)}{\epsilon_{+}(\mathbf{p})}\left[f\left(\epsilon_{+}(\mathbf{p})\right)-f\left(-\epsilon_{+}(\mathbf{p})\right)\right].
 \end{equation}
The contribution of this term into optical conductivity does not depend on the frequency and we neglect it in our studies.

To evaluate the real parts of longitudinal optical conductivity along the x- and y-directions, we first calculate traces with time-dependent velocity operators, which are obtained from Eqs.\eqref{eq:Heisenberg-semi-1} and \eqref{eq:solution-matrix-semi-Dirac},
\begin{align}
		&\Tr[e^{-i H_{semi}s}v_x(t)v_x(0)]= \int \frac{d^2 p}{(2\pi)^2} \frac{8 a^2 p_x^2 \left(v^2 p_y^2 \cos \left((s-2 t) \epsilon_{+}\right)+\left(a p_x^2+\Delta
		\right)^2 \cos \left(s \epsilon_{+}\right)\right)}{\epsilon_{+}^2},\\
	&\Tr[e^{-i H_{semi}s}v_y(t)v_y(0)]=\int \frac{d^2 p}{(2\pi)^2} \frac{2 v^2 \left[\left(a p_x^2+\Delta \right)^2 \cos \left((s-2 t)
		\epsilon_{+}\right)+ v^2
		p_y^2 \cos \left(s\epsilon_{+}\right)\right]}{\epsilon_{+}^2},\quad \epsilon_{+}\equiv \epsilon_{+}(\mathbf{p}).
\end{align}
As described in the main text, we then make Fourier transforms over $t$ and $s$ to obtain the delta-functions under integrals which technically simplify integrals. The resulting expressions for longitudinal optical conductivity are:
\begin{align}
	\label{eq:Re-xx-integral}
	&\Re\sigma_{xx}(\omega)=\frac{2 e^2}{\omega}\int_{-\infty}^{\infty} \frac{d E}{2\pi} [f(E)-f(E+\omega)]\int d^2p \frac{a^2 p_x^2}{\epsilon_{+}^2} \times\nn
	&\times\left[ \delta \left(E+\epsilon_{+}\right) \left(v^2 p_y^2 \delta \left(\omega
		-2\epsilon_{+}\right)+\delta
		(\omega ) \left(a p_x^2+\Delta \right)^2\right)+ \delta
		\left(E-\epsilon_{+}\right)
		\left(v^2 p_y^2 \delta \left(\omega +2\epsilon_{+}\right)+\delta (\omega ) \left(a p_x^2+\Delta
		\right)^2\right)\right],\\
		\label{eq:Re-yy-integral}
	&\operatorname{Re} \sigma_{yy}(\omega)=\frac{e^{2}}{2 \omega} \int_{-\infty}^{\infty} \frac{d E}{2\pi} [f(E)-f(E+\omega)]\int d^{2} p
\frac{v^2}{\epsilon_{+}^2} \times\nn
	&\times\left[ \delta \left(E+\epsilon_{+}\right) \left(\left(a p_x^2+\Delta
		\right){}^2 \delta \left(\omega -2 \epsilon_{+}\right)+v^2 \delta
		(\omega ) p_y^2\right)+\delta \left(E-\epsilon_{+}\right) \left(\left(a p_x^2+\Delta
		\right){}^2 \delta \left(\omega +2 \epsilon_{+}\right)+v^2 \delta
		(\omega ) p_y^2\right)\right].
\end{align}
To perform the integration over momentum, we use the symmetry $p_x\to -p_x$, $p_y\to -p_y$ of the integrals  and the following change of coordinates that simplifies square root in $\epsilon_{+}$:
\begin{align}
	a p_{x}^{2}+\Delta=L \cos \phi, \quad v p_{y}=L \sin \phi,\quad \epsilon_{+}=L.
\label{variables-change}
\end{align}
For the functions even in $p_x$ and $p_y$ we can write
\be
\int d^2p f(p_x,p_y)=4\int\limits_0^\infty d p_x d p_y f(p_x,p_y)=\int\limits_0^\infty dL \int\limits_0^\pi d\phi \frac{2L\,\theta(L\cos\phi-\Delta)}{v\sqrt{a(L\cos\phi-\Delta)}} f\left(\sqrt{\frac{L\cos\phi-\Delta}{a}},\frac{L\sin\phi}{v}\right).
\label{momentum-integration}
\ee
The presence of the theta function takes into account  that the regions of integration of the $L$ and $\phi$ variables will be different depending on the sign of the $\Delta$ parameter.  In what follows, we extensively use the following integral  (Eq. 3.197.8 from book \cite{Gradstein_Ryzhik}):
\begin{align}\label{eq:conductivity-gradstein-integral}
	\int_{0}^{u} x^{\nu-1}(x+a)^{\lambda}(u-x)^{\mu-1} d x=a^{\lambda} u^{\mu+\nu-1} \mathrm{~B}(\mu, \nu)\,{}_{2} F_{1}\left(-\lambda, \nu ; \mu+\nu;
-\frac{u}{a}\right),\quad\arg\frac{u}{a}<\pi.
\end{align}
Performing the momentum integration in Eqs.\eqref{eq:Re-xx-integral}, \eqref{eq:Re-yy-integral} by means of Eq .(\ref{momentum-integration}),
we obtain:
\begin{align}\label{integrals-L,phi-xx}
	&\text{xx}:\quad\int d^2 p[\dots]=\frac{2\sqrt{a}}{v}\int\limits_{0}^{\infty}dL\,\int\limits_{0}^{\pi}d\phi L\sqrt{(L\cos\phi-\Delta)}
\theta(L\cos\phi-\Delta) \times\nn
	&\left[ \delta \left(E+L\right) \left(\sin^2\phi \delta \left(\omega-2L\right)+\delta(\omega ) \cos^2\phi\right)+ \delta
	\left(E-L\right)\left(\sin^2\phi \delta \left(\omega +2L\right)+\delta (\omega ) \cos^2\phi\right)\right],\\
	&\text{yy}:\quad\int d^2 p[\dots]=\frac{2v}{\sqrt{a}} \int\limits_{0}^{\infty}dL\,\int\limits_{0}^{\pi}\frac{L\,d\phi}{\sqrt{L\cos\phi-\Delta}} \theta(L\cos\phi-\Delta)\times\nn
	&\left[ \cos^2\phi(\delta \left(E+L\right)  \delta \left(\omega-2L\right)+\delta(E-L)\delta(\omega+2L))+\sin^2\phi\delta
	(\omega ) (\delta(E+L)+\delta\left(E-L\right))
	\right].
\label{integrals-L,phi-yy}
\end{align}
The integration over angle depends on the sign of $\Delta$. For $1>\delta=\Delta/L\ge0$, we find the following four integrals:
\begin{align}
	I_1^{xx}(\delta)&=\int_{0}^{\phi_L}\sqrt{\cos\phi-\delta}\sin^2\phi\,d\phi=\frac{2\sqrt{2}}{15}\left[2(3+\delta^2)E(k)-(3+\delta)(1+\delta)K(k)\right],\\
I_2^{xx}(\delta)&=\int_{0}^{\phi_L}\sqrt{\cos\phi-\delta}\cos^2\phi\,d\phi =\frac{\sqrt{2}}{15}  \left[(1+\delta) (2 \delta -9) K(k)+\left(18-4 \delta ^2\right) E(k)\right],\\
I_{1}^{yy}(\delta)&=\int\limits_{0}^{\phi_L}\frac{\sin^2\phi d\phi}{\sqrt{\cos\phi-\delta}}=\frac{2\sqrt{2}}{3}\left[(1+\delta)K(k)-2\delta E(k)\right],\\
I_{2}^{yy}(\delta)&=\int\limits_{0}^{\phi_L}\frac{\cos^2\phi d\phi}{\sqrt{\cos\phi-\delta}}=\frac{\sqrt{2}}{3}\left[(1-2\delta)K(k)+4\delta E(k)\right],
\end{align}
where $K(k)$ and $E(k)$ are complete elliptic integrals, $k=\sqrt{\frac{1-\delta}{2}}$, and $\phi_L=\arccos(\delta)$.
To calculate the above integrals we made the variable change $x=\cos\phi$, then used Eq.(\ref{eq:conductivity-gradstein-integral}), the relation
\be
{}_2F_1(a,b;c;z)=(1-z)^{-a}{}_2F_1\left(a,c-b;c;\frac{z}{z-1}\right).
\ee
and Eqs. 7.3.2.18, 7.3.2.20 and 7.3.2.75 from the book \cite{Prudnikov3}.

Case $\Delta<0$: in this case the angular integration is separated into two regions,
	\begin{align}
		 \phi \in \begin{cases}{\left[0, \arccos \frac{-|\Delta|}{L}\right],} & L>|\Delta|, \\ {[0, \pi],} & L \leq|\Delta|.\end{cases}
	\end{align}
This example can be seen as integrating with the centers in the Dirac point.
Performing integration over angle in Eqs.(\ref{integrals-L,phi-xx}), (\ref{integrals-L,phi-yy}) we find the following: the integrals for $L>|\Delta|$ are the same as in $\Delta>0$ case with the changes $\Delta\to-|\Delta|$. The integrals for $L<|\Delta|$ ($|\delta|>1$) are different and have the following form:
\begin{align}
	I_{3}^{xx}(\delta<-1)&=\int_{0}^{\pi}\sqrt{\cos\phi+|\delta|}\sin^2\phi\, d\phi
=\frac{4}{15}\sqrt{|\delta|+1}\left[(3+\delta^2)E(k')-|\delta|(|\delta|-1)K(k')\right],\\
	I_{4}^{xx}(\delta<-1)&=\int_{0}^{\pi}\sqrt{\cos\phi+|\delta|}\cos^2\phi\, d\phi=\frac{2}{15}\sqrt{|\delta|+1}\left[(9-2\delta^2)E(k')
+2|\delta|(|\delta|-1)K(k')\right],\\
I_{3}^{yy}(\delta<-1)&=\int_{0}^{\pi}\frac{\sin^2\phi\, d\phi}{\sqrt{\cos\phi+|\delta|}}
=\frac{4}{3}\sqrt{|\delta|+1}\left[|\delta|E(k')-(|\delta|-1)K(k')\right],\\
I_{4}^{yy}(\delta<-1)&=\int_{0}^{\pi}\frac{\cos^2\phi\, d\phi}{\sqrt{\cos \varphi+|\delta|}}=\frac{2}{3\sqrt{|\delta|+1}}
\left[-2|\delta|(|\delta|+1)E(k')+(1+2\delta^2)K(k')\right],
\end{align}
where $k'=\sqrt{\frac{2}{|\delta|+1}}$.

Evaluating the integrals over $L$ in all these cases gives the following results for longitudinal conductivities in the $x-$ and $y-$directions:
\begin{align}
	&\operatorname{Re} \sigma_{x x}(\omega)= \frac{e^{2}}{4 \pi\hbar \omega} \int_{-\infty}^{\infty} d E[f(E)-f(E+\omega)]
\frac{4|E|^{3 / 2} a^{1 / 2}}{v}\times\nn
	&\times\begin{cases}
		\begin{array}{c}
			2\Theta(|\Delta|-|E|) \left(I_{3}^{xx}(\Delta/|E|) \delta \left(\omega
			+2 E \right)+I_{4}^{xx}(\Delta/|E|)\delta
			(\omega ) \right)+\\
			+2\Theta(|E|-|\Delta|) \left(I_{1}^{xx}(\Delta/|E|) \delta \left(\omega
			+2 E\right)+I_{2}^{xx}(\Delta/|E|)\delta
			(\omega )\right)
		\end{array} ,&\Delta<0,\\
			& \\
		\frac{8\pi^{3 / 2}}{5 \sqrt{2} \Gamma^{2}\left(\frac{1}{4}\right)}[2 \delta(\omega+2 E)+3 \delta(\omega)],&\Delta=0,\\
				& \\
		2\Theta(|E|-\Delta)\left[I_{1}^{xx}(\Delta/|E|) \delta(\omega+2 E)+I_{2}^{xx}(\Delta/|E|) \delta(\omega)\right],&\Delta>0,	
	\end{cases}
\end{align}
and
\begin{align}
	\operatorname{Re} \sigma_{yy}(\omega)&=\frac{e^{2}}{4\pi \hbar \omega} \int_{-\infty}^{\infty} d E[f(E)-f(E+\omega)] \frac{v\sqrt{|E|}}{\sqrt{a}}
\times\nn
	&\times\begin{cases}
		\begin{array}{c}
			2\Theta(|\Delta|-|E|)\left(I_{4}^{y y}(\Delta/|E|) \delta(\omega+2 E)+I_{3}^{y y}(\Delta/|E|) \delta(\omega)\right)+\\
			+2\Theta(|E|-|\Delta|)\left(I_{2}^{y y}(\Delta/|E|) \delta(\omega+2 E)+I_{1}^{y y}(\Delta/|E|) \delta(\omega)\right)
		\end{array},&\Delta<0,\\
		& \\
		 \frac{\Gamma^{2}\left(\frac{1}{4}\right)}{3 \sqrt{2\pi}}
		[\delta(\omega+2 E)+2 \delta(\omega)],&\Delta=0,\\
		& \\
		2\Theta(|E|-\Delta)
		\bigg[I_{2}^{yy}(\Delta/|E|) \delta \left(\omega+2E\right)+I_{1}^{yy}(\Delta/|E|)
		\delta(\omega ) \bigg],&\Delta>0.
	\end{cases}
\end{align}
Separating interband ac and intraband dc parts, we find the results given by Eqs.\eqref{eq:Re-sigma-semi-xx-AC} and \eqref{eq:Re-sigma-semi-xx-DC} together with \eqref{eq:Re-sigma-semi-yy-AC} and \eqref{eq:Re-sigma-semi-yy-DC} in the main text.

    \section{Longitudinal conductivity of the gapped dice model.}

    First we evaluate traces of commutators with matrix exponential of the Hamiltonian:
    \begin{align}
    	&\Tr[e^{-iHs}v_x(t)v_x(0)]=\frac{v_F^2 \cos \left(p s v_F\right)
    		\left(2\left(p_y^2+p_z^2\right)
    		p^2 \cos \left(p t
    		v_F\right)+4p_x^2 p^2\right)}{2 p^4}+\nn
    	&+\frac{v_F^2\left(2
    		\left(p_y^2+p_z^2\right) \left(p^2 \sin \left(p s
    		v_F\right) \sin \left(p t
    		v_F\right)+p^2 \cos \left(p
    		t v_F\right)\right)\right)}{2 p^4},\\
    	&\Tr[e^{-iHs}v_y(t)v_y(0)]=\frac{v_F^2 \left(\cos \left(p s v_F\right)
    		\left(2\left(p_x^2+p_z^2\right)
    		p^2 \cos \left(p t
    		v_F\right)+4p_y^2 p^2\right)\right)}{2 p^4}+\nn
    	&+\frac{v_F^2\left(+2
    		\left(p_x^2+p_z^2\right) \left(p^2 \sin \left(p s
    		v_F\right) \sin \left(p t
    		v_F\right)+p^2 \cos \left(p
    		t v_F\right)\right)\right)}{2 p^4}.
    \end{align}
    Next, we Fourier transform this expressions twice with respect to $t\to \omega$ and $s\to E$, and integrate over the  polar angle
    \begin{align}
    	\mathcal{F}_{t,s}\Tr[e^{-iHs}v_x(t)v_x(0)]&=\delta (E) \left(\frac{\pi  v_F^2 \left(p^2+
    		p_z^2\right) \delta \left(\omega -p v_F\right)}{2
    		p^2}+\frac{\pi  v_F^2 \left(p^2+ p_z^2\right) \delta
    		\left(\omega +p v_F\right)}{2 p^2}\right)+\nn
    	&+\delta
    	\left(E+p v_F\right) \left(\frac{\pi  v_F^2
    		\left(p^2+ p_z^2\right) \delta \left(\omega -p
    		v_F\right)}{2 p^2}+\frac{\pi  (p^2-p_z^2)
    		v_F^2\delta (\omega )}{p^2}\right)+\nn
    	&+\delta \left(E-p v_F\right)
    	\left(\frac{\pi  v_F^2 \left(p^2+ p_z^2\right) \delta
    		\left(\omega +p v_F\right)}{2 p^2}+\frac{\pi  (p^2-p_z^2)
    		v_F^2\delta (\omega )}{p^2}\right).
    \end{align}
Due to isotropy of the model we get the same result for the Fourier transform $\mathcal{F}_{t,s}\Tr[e^{-iHs}v_y(t)v_y(0)]$.   	

The longitudinal conductivity is given by the expression
    \begin{align}
    	&\Re \sigma_{xx}(\omega)=\frac{\pi e^{2} }{\omega} \int_{-\infty}^{\infty} d E [f(E)-f(E+ \omega)]\int_{0}^{\infty}\frac{k \,dk}{(2\pi)^2}
    \mathcal{F}_{t,s}\Tr[e^{-iHs}v_x(t)v_x(0)].
    \end{align}
where $k=\sqrt{p_x^2+p_y^2}$.
 Finally, performing integrations we find
    \begin{align}
    	&\Re \sigma_{xx}(\omega)=\frac{e^2}{4}\bigg[\delta(\omega)\int_{-\infty}^{\infty} d E\frac{f(E)-f(E+\omega)}{\omega} \Theta\left(|E|-v_{F}p_z\right) \frac{|E|^2-v_F^2 p_z^2}{|E|}+\bigg.\nn
    	&+\frac{f(-\omega)-f(\omega)}{\omega}\left(\frac{1}{2}+\frac{p_z^2 v_F^2}{2\omega^2}\right)
    \left[\omega\Theta(\omega-\Delta)+|\omega|\Theta(-\omega-\Delta)\right]\bigg.\bigg]\nn
    	&=\frac{ e^2 }{4}\bigg[\delta(\omega)\int_{-\infty}^{\infty} d E\frac{f(E)-f(E+\omega)}{\omega} \Theta\left(|E|-\Delta\right)
    \frac{|E|^2-\Delta^2}{|E|}+\frac{f(-\omega)-f(\omega)}{\omega}\frac{\omega^2+\Delta^2}{2|\omega|}\Theta(|\omega|-\Delta)\bigg],
    \end{align}
 where in the last equality we took into account that $v_Fp_z=\Delta>0$.
This expression appears in the main text, Eq.(\ref{eq:re-xx-dice}), in slightly different form and is plotted for different values of parameters.

		\section{Evalution of Hall conductivity $\sigma_{xy}$ in gapped dice model}
	Let us evaluate the quasiparticle velocity operator averages for the Hall conductivity. First, we evaluate the matrix traces:
	\begin{align}
		&\operatorname{tr}\left[e^{-i v_{F} \mathbf{S p} s}\left(v_{x}(t) v_{y}(0)+v_{y}(t) v_{x}(t)\right)\right]=-\frac{2 v_F^2 p_x p_y \left(\cos \left(p v_F (s-t)\right)-2 \cos \left(p s v_F\right)+\cos
			\left(p t v_F\right)\right)}{p^2},\\
		&\operatorname{tr}\left[e^{-i v_{F} \mathbf{S p} s} \left(v_{x}(t) v_{y}(0)-v_{y}(t) v_{x}(0)\right)\right]=\frac{2 v_F^2 p_z \left(\sin \left(p v_F (s-t)\right)-\sin \left(p t v_F\right)\right)}{p}.
	\end{align}
The first trace vanishes after the angle integration. Thus the symmetric part is absent for the Hall conductivity, as expected. For the antisymmetric part we find (again $k=\sqrt{p_x^2+p_y^2}$):
\begin{align}
&\Tr[\delta(E-H)\left(v_{x}(t) v_{y}(0)-v_{y}(t) v_{x}(0)\right)]=\frac{V}{2 \pi} \int_{-\infty}^{\infty} d s e^{i E s} \int_{0}^{\infty}
\frac{k d k}{(2 \pi)}\frac{2 v_F^2 p_z \left(\sin \left(p v_F (s-t)\right)-\sin \left(p t v_F\right)\right)}{p}=\nn
&=V \int_{0}^{\infty} \frac{k d k}{(2 \pi)}\frac{2 v_F^2 p_z}{p} \left(\frac{e^{-i p v_F t}\delta(E+p v_F)-e^{i p v_F t}\delta(E-p v_F)}{2i}-
\delta(E)\sin \left(p t v_F\right)\right).
	\end{align}
Next we perform integration over time and find
\begin{align}
&\int_{-\infty}^{\infty} d t e^{i \omega t} \Tr[\delta(E-H)\left(v_{x}(t) v_{y}(0)-v_{y}(t) v_{x}(0)\right)]=\nn
&=V \int_{0}^{\infty} k d k\frac{2 v_F^2 p_z}{p} \left(\frac{\delta(\omega- p v_F )\delta(E+p v_F)-\delta(\omega+p v_F )\delta(E-p v_F)}{2i}
-\delta(E)\frac{\delta(\omega+p v_F )-\delta(\omega-p v_F )}{2i}\right).
	\end{align}
 Thus, for the imaginary part of the Hall conductivity we find
	\begin{align}
\operatorname{Im} \sigma_{[x, y]}(\omega)&=\frac{1}{2}\frac{e^{2}}{4 \hbar \omega} \int_{0}^{\infty} k d k\frac{2 v_F^2 p_z}{p} \int_{-\infty}^{\infty} d E [f(E)-f(E+\hbar \omega)]\left(-\delta(\omega- p v_F )\delta(E+p v_F)+\delta(\omega+p v_F )\delta(E-p v_F)+\right.\nn
&+\left.\delta(E)[\delta(\omega+p v_F )-\delta(\omega-p v_F )]\right)=\nn
&=\frac{e^{2}v_F^2 p_z}{4 \hbar \omega} \int_{0}^{\infty} \frac{k d k }{p}\bigg(\delta(\omega+p v_F)[f(pv_F)-f(pv_F+\omega)+f(0)-f(\omega)]-
\bigg.\nn
&-\bigg.\delta(\omega- p v_F )[f(-pv_F)-f(-pv_F+\omega)+f(0)-f(\omega)]\bigg).
	\end{align}
	Also in the first line we canceled $\rho(E)$ and $V$ with the normalization $\Tr\delta(E-H)$. The factor $1/2$ in the first line of the last
equation accounts for the definition of the antisymmetric part of the tensor. Now we can integrate over momenta and obtain
	\begin{align}
		&\operatorname{Im} \sigma_{[x, y]}(\omega>0)=\frac{e^{2}}{4 \omega} v_F p_z \Theta\left(\omega-v_{F} |p_{z}|\right)(f(\omega)-f(-\omega)),\\
		&\operatorname{Im} \sigma_{[x, y]}(\omega<0)=\frac{e^{2}}{4 \omega}  v_F p_z\Theta\left(-\omega-v_{F}| p_{z}|\right)(f(-\omega)-f(\omega)).
	\end{align}
	Combining these formulas together we arrive at Eq.(\ref{eq:Hall-conductivity-dice}).
	
Now using the Kramers-Kronig relation we can evaluate the real part:
	\begin{align}\label{eq:integral-Hall-Re-dice}
		\operatorname{Re} \sigma_{[x, y]}(\Omega)=\frac{1}{\pi} \text { P.v. } \int\limits_{-\infty}^{\infty} \frac{d \omega \operatorname{Im} \sigma_{[\mu, \nu]}(\omega)}{\omega-\Omega}=\frac{ e^2v_F p_z}{4\pi} \text { P.v. } \int\limits_{-\infty}^{\infty} d\omega\frac{\Theta\left(|\omega|-v_{F} |p_{z}|\right)
(f(|\omega|)-f(-|\omega|))}{\omega(\omega-\Omega)}.
	\end{align}
It is easy to check that $\operatorname{Re} \sigma_{[x, y]}(\Omega)$ is even function in $\Omega$ by changing the integration variable. The integral simplifies for the zero temperature when
\begin{equation}
	f\left(| \omega|\right)-f\left(-| \omega|\right) \rightarrow \theta\left(\mu-|\omega|\right)-\theta\left(| \omega|+\mu\right)=-\theta(| \omega|-|\mu|).
\end{equation}
Thus, Eq.\eqref{eq:integral-Hall-Re-dice} gives Eq.(\ref{Hall-dice-T0}).

	\section{Conductivities of the Lieb model.}
	\label{appendix:Lieb-conductivity-details}
The system of equations for the Gell-Mann matrices is given by Eq.(\ref{eq:lambda-matrices}) with the initial values $\lambda_i(t=0)=\lambda_i$.
	The solutions for the $v_x(t)$ and $v_y(t)$ are defined as vectors in the Gell-Mann basis (the identity matrix is not present because it does not evolve with time and the coefficient before this matrix is zero): $v_x(t)=v_F\left(e^{A t}\right)_{1j}\lambda_j$, $v_y(t)=v_F\left(e^{A t}\right)_{6j}\lambda_j$ where
	\begin{align}\label{eq:velocity-x-Gell-Mann}
		&\left(e^{A t}\right)_{1j}= \left(
		\begin{array}{c}
			\frac{\Delta ^2 q_x^2 \cos \left(2 p t v_F\right)+p q_y^2 \left(p \cos \left(p t v_F\right) \cos \left(\Delta  t v_F\right)-\Delta  \sin \left(p t v_F\right) \sin \left(\Delta  t v_F\right)\right)+(p^2-\Delta^2 ) q_x^2}{p^2(p^2-\Delta^2 ) } \\
			-\frac{\cos \left(p t v_F\right) \left(2 \Delta  q_x^2 \sin \left(p t v_F\right)+p q_y^2 \sin \left(\Delta  t v_F\right)\right)+\Delta  q_y^2 \sin \left(p t v_F\right) \cos \left(\Delta  t v_F\right)}{p(p^2-\Delta^2)} \\
			\frac{q_x \sin \left(p t v_F\right) \left(\Delta  \left(2 q_x^2+q_y^2\right) \sin \left(p t v_F\right)+p q_y^2 \sin \left(\Delta  t v_F\right)\right)}{p^2 (p^2-\Delta^2 ) } \\
			\frac{q_y \sin \left(p t v_F\right) \left(2 \Delta  q_x^2 \sin \left(p t v_F\right)+p \left(q_y^2-q_x^2\right) \sin \left(\Delta  t v_F\right)\right)}{p^2 (p^2-\Delta^2 ) } \\
			\frac{q_y \sin \left(p t v_F\right) \cos \left(\Delta  t v_F\right)}{p} \\
			\frac{q_x q_y \left(-\Delta ^2-p^2 \cos \left(p t v_F\right) \cos \left(\Delta  t v_F\right)+\Delta ^2 \cos \left(2 p t v_F\right)+\Delta  p \sin \left(p t v_F\right) \sin \left(\Delta  t v_F\right)+p^2\right)}{p^2(p^2-\Delta^2 ) } \\
			-\frac{q_x q_y \left(-\Delta  \sin \left(2 p t v_F\right)+\Delta  \sin \left(p t v_F\right) \cos \left(\Delta  t v_F\right)+p \cos \left(p t v_F\right) \sin \left(\Delta  t v_F\right)\right)}{p(p^2-\Delta^2)} \\
\frac{\sqrt{3} q_x q_y^2 \sin \left(p t v_F\right) \left(p \sin \left(\Delta  t v_F\right)-\Delta  \sin \left(p t v_F\right)\right)}{p^2 (p^2-\Delta^2 ) } \\
		\end{array}
		\right)^T,
	\end{align}
	\begin{align}\label{eq:velocity-y-Gell-Mann}
		&\left(e^{A t}\right)_{6j}=\left(
		\begin{array}{c}
			\frac{q_x q_y \left(-\Delta ^2-p^2\cos \left(p t v_F\right) \cos \left(\Delta  t v_F\right)+\Delta ^2 \cos \left(2 p t v_F\right)+\Delta  p
\sin \left(p t v_F\right) \sin \left(\Delta  t v_F\right)+p^2\right)}{p^2(p^2-\Delta^2 ) } \\
			\frac{q_x q_y \left(-\Delta  \sin \left(2 p t v_F\right)+\Delta  \sin \left(p t v_F\right) \cos \left(\Delta  t v_F\right)+p \cos\left(p t v_F\right) \sin \left(\Delta  t v_F\right)\right)}{p(p^2-\Delta^2)} \\
			\frac{q_y \sin \left(p t v_F\right) \left(\Delta  \left(2 q_x^2+q_y^2\right) \sin \left(p t v_F\right)-p q_x^2 \sin\left(\Delta  t v_F\right)\right)}{p^2 (p^2-\Delta^2 ) } \\
			\frac{q_x \sin \left(p t v_F\right) \left(p \left(q_x^2-q_y^2\right) \sin \left(\Delta  t v_F\right)+2 \Delta  q_y^2
\sin\left(p t v_F\right)\right)}{p^2 (p^2-\Delta^2 ) } \\
			-\frac{q_x \sin \left(p t v_F\right) \cos \left(\Delta  t v_F\right)}{p} \\
			\frac{p q_x^2 \left(p \cos \left(p t v_F\right) \cos \left(\Delta  t v_F\right)-\Delta  \sin \left(p t v_F\right) \sin \left(\Delta  t v_F\right)\right)+\Delta ^2 q_y^2 \cos \left(2 p t v_F\right)+(p^2-\Delta^2 ) q_y^2}{p^2
				(p^2-\Delta^2 ) } \\
			\frac{\Delta  q_x^2 \sin \left(p t v_F\right) \cos \left(\Delta  t v_F\right)+p q_x^2 \cos \left(p t v_F\right) \sin \left(\Delta  t v_F\right)+\Delta  q_y^2 \sin \left(2 p t v_F\right)}{p(p^2-\Delta^2)} \\
-\frac{\sqrt{3} q_y \sin \left(p t v_F\right) \left(p q_x^2 \sin \left(\Delta  t v_F\right)+\Delta  q_y^2 \sin \left(p t v_F\right)\right)}{p^2 (p^2-\Delta^2 ) } \\
\end{array}\right)^T.
	\end{align}
	
Integrating over $t$ and $s$ in Eqs.(\ref{eq:Re-sigma-mu-nu}), (\ref{eq:trace-Fourier-sE-general}) we find:
	\begin{align}
		\Re \sigma_{xx}(\omega)&=\addDO{2\pi}\frac{\pi e^{2} v_F^2}{2 \omega} \int_{-\infty}^{\infty} d E [f(E)-f(E+ \omega)]\int_{0}^{\infty}\frac{k \,dk}{(2\pi)^2}\times\nn
		&\times\left[\delta \left(E-p v_F\right) \left(\frac{ \Delta ^2  \delta \left(\omega +2 p v_F\right)}{p^2}+\frac{ \delta (\omega )
			(p^2-\Delta^2 ) }{p^2}-\left(\frac{  \Delta  }{2 p}-\frac{1}{2} \right) \delta
		\left(\omega +(p -\Delta)  v_F\right)\right)+\right.\nn
		&+\left.\delta \left(E+p v_F\right) \left(\frac{  \Delta ^2 \delta \left(\omega -2 p
			v_F\right)}{p^2}+\frac{ \delta (\omega ) (p^2-\Delta^2 ) }{p^2}+\frac{ (\Delta +p) \delta \left((p+\Delta )
			v_F-\omega \right)}{2 p}\right)+\right.\nn
		&+\left.\delta \left(E-\Delta  v_F\right) \left(\left(-\frac{  \Delta   }{2 p}+\frac{1}{2}\right)  \delta \left(\omega -(p -\Delta  )v_F\right)+\left(\frac{ \Delta   }{2 p}+\frac{1}{2} \right) \delta \left(\omega +(p+\Delta ) v_F\right)\right)\right],
	\end{align}
where $k=\sqrt{q_x^2+q_y^2}$.
	At the same time we find $\Im\sigma_{[x,y]}=0$ after taking the trace of the product of velocities.
	Next, we calculate the integrals which involve the delta-functions, first we integrate over $E$ and then over momenta, we get the expression
\begin{align}
	&\Re\sigma_{xx}(\omega)=\frac{e^2 }{4}\left[\delta(\omega)\int\limits_{\Delta v_F}^{\infty}pv_F d(p v_F)\left(\frac{1}{4T\cosh^2((p v_F-\mu)/2T)}
+\frac{1}{4T\cosh^2((p v_F+\mu)/2T)}\right)\frac{p^{2}-\Delta^{2}}{p^{2}}+\right.\nn
	&\left.+\Theta(|\omega|-2\Delta v_F)\left[\frac{2\Delta^2 v_F^2}{ \omega^2}\left(f\left(-\frac{|\omega|}{2}\right)-f\left(\frac{|\omega|}{2}\right)\right)+\frac{1}{2}(f(\Delta v_F-|\omega|)-f(\Delta v_F))\right]+\frac{f(\Delta v_F)-f(\Delta v_F+ |\omega|)}{2} \right],
\end{align}
which is in fact Eq.(\ref{eq:re-sigma-Lieb}) in the main text after restoring $\hbar$. The remaining integral can be evaluated in terms of the polylogarithm functions.

	\bibliography{dice_bib}
	
\end{document}